\documentclass[useAMS,usenatbib,usegraphicx]{mn2e}

\usepackage{amsmath}
\usepackage{amssymb}
\usepackage{txfonts}
\usepackage[colorlinks=true,citecolor=blue,linkcolor=blue,urlcolor=blue]{hyperref}
\usepackage[usenames,dvipsnames]{xcolor}
\newcommand{\dd}{\rmn{d}}
\newcommand{\tens}[1]{\mathbfss{#1}}
\renewcommand{\pi}{\upi}

\DeclareMathOperator{\tr}{tr}
\DeclareMathOperator{\erf}{erf}

\graphicspath{{./figures/}}

\begin{document} 

\title[Ellipticity and prolaticity at density maxima]{Ellipticity and prolaticity of the initial gravitational-shear field at the position of density maxima}

\author[C.~Angrick]{C.~Angrick\thanks{eMail: angrick@uni-heidelberg.de}\\
Zentrum f\"ur Astronomie, Institut f\"ur Theoretische Astrophysik, Universit\"at Heidelberg, Philosophenweg~12, D-69120~Heidelberg, Germany}

\maketitle

\begin{abstract}
Dark-matter haloes are supposed to form at the positions of maxima in the initial matter density field. The gravitational-shear field's ellipticity and prolaticity that serve as input for the ellipsoidal-collapse model, however, are derived from a distribution that does not take the additional maximum constraint into account. In this article, I quantify the variations of the most probable and the expected values of the ellipticity and the prolaticity when considering this additional constraint as well as the implications for the ellipsoidal-collapse model. Based on the statistics of Gaussian random fields, it is possible to set up a joint distribution for the eigenvalues of the gravitational-shear tensor and the matter density that incorporates the maximum constraint by invoking a vanishing first derivative and a negative definite second derivative of the density field into the calculation. In the density range relevant for cosmological structure formation, both the most probable and the expected value of the ellipticity calculated from the standard distribution used in the literature are about 3--8 per cent higher compared to the ones calculated under the additional assumption of a density maximum. Additionally, the analogous quantities for the prolaticity do not vanish but acquire slightly positive values in the range of $10^{-3}$--$10^{-2}$. For large overdensities, the predictions from both distributions converge. The values for $\delta_\rmn{c}$ and $\Delta_\rmn{v}$ derived from the ellipsoidal-collapse model using the standard distribution for the initial ellipticity and prolaticity are up to 4 and 6 per cent higher, respectively, than those obtained taking the additional maximum constraint into account in the range of $10^{13}$--$10^{15}~h^{-1}~\rmn{M}_\odot$ in mass and 0--2 in redshift.
\end{abstract}


\section{Introduction}
\label{sec:introduction}

Galaxies and galaxy clusters as collapsed objects are supposed to form from peaks in the initial matter density field. After an overdense patch has decoupled from the Hubble expansion, the extent of this patch is supposed to reach a maximum value, then to shrink again until the collapse stops when the structure reaches virial equilibrium. Analytical models for this evolution include the spherical-collapse model (e.g.\ \citealt{Gunn1972,Lahav1991,Lacey1993,Wang1998}; \citealt*{Pace2010}) and the ellipsoidal-collapse model (e.g.\ \citealt{Icke1973,White1979,Barrow1981}; \citealt*{Bartelmann1993}; \citealt{Eisenstein1995,Bond1996}).

Most of the theories of inflation predict that the density field is initially a random field whose non-Gaussianity is small \citep{Planck2013}. Since also derivatives and integrals of Gaussian random fields are again Gaussian, a multivariate Gaussian random field can be used to derive the initial number density of peaks in the density field as a function of peak height by applying respective constraints to the density field's first and second derivatives \citep{Bardeen1986}.

Not only can the density field be described as a Gaussian random field, but through Poisson's equation, also the gravitational field has to follow a Gaussian distribution and therefore also the eigenvalues of the gravitational-shear tensor can be derived from the statistics of Gaussian random fields \citep{Doroshkevich1970,Zeldovich1970}. These eigenvalues can be combined to define the ellipticity and the prolaticity of the gravitational-shear field whose distribution for a given value of the overdensity can be derived by transformation of variables (\citealt{Bardeen1986,Weygaert1996}; \citealt*{Sheth2001}) and either their most probable or their expected values are used to set the initial conditions in the ellipsoidal-collapse model by \citet{Bond1996}.

However, this distribution does not take into account that haloes should form at \emph{maxima} in the density field, but rather yields the probability density for ellipticity and prolaticity for \emph{any} overdensity irrespective if it is taken at a maximum or not. Recently, attempts have been made to take the maximum constraint into account when calculating the gravitational-shear tensor's conditional eigenvalues for given eigenvalues of the density's Hessian \citep{Rossi2012} as well as the conditional ellipticity and prolaticity from them \citep{Rossi2013} by extending the formalism of \citet{Doroshkevich1970} and \citet{Sheth2001}. \citet{Rossi2012,Rossi2013}, however, only covers one of the two constraints necessary for a density maximum, i.e.\ that the density's Hessian has to be negative definite. The other necessary condition, namely the vanishing first derivative of the density, is not taken into account. Additionally, \citet{Rossi2013} does not derive the distribution for the unconstrained ellipticity and prolaticity necessary for the ellipsoidal-collapse model by \citet{Bond1996}.

Comparisons to numerical simulations, however, reveal that the simple picture of this model is not fully realized. For example, \citet{Ludlow2011a} found out that not all collapsed haloes form from a density peak in the initial Gaussian random field whose characteristic mass is close to that of the halo. Only $\sim$70 per cent of well-resolved haloes in their simulation possess this preference, whereby heavier haloes have the tendency to stay near the position of the initial peak, while low-mass haloes increasingly tend to move away from the location of the initial density peak due to the gravitational influence of the surrounding structure. Also \citet*{Elia2012} claim that relating virialized haloes to density peaks of a given height is too simplistic to forecast the density bias parameters correctly with the analytic model by \citet{Desjacques2010}.

The works by both \citet*{Ludlow2011b} and \citet*{Despali2013} have shown that the initial protohaloes are not spherical as assumed in the model by \citet{Bond1996} but have already a rather ellipsoidal shape. Additionally, they found out that for most haloes the deformation and the mass tensors are all well aligned and that the longest axis corresponds to the direction of maximal compression. While the first result corresponds to one of the main assumptions of the ellipsoidal-collapse model, the latter is in direct contrast with it. Accounting for these deviations in the analytical model results in lowering the critical overdensity $\delta_\rmn{c}$ significantly for a given mass \citep{Ludlow2011b} and in a possible inversion of axes so that the initially shortest axis might become the longest during the evolution and vice versa \citep{Ludlow2011b,Despali2013}.

Despite the differences between results from numerical simulations and the simple ellipsoidal-collapse model by \citet{Bond1996}, we will focus on the latter taking only into account modifications by \citet{Angrick2010} in the following to clarify if incorporating the maximum constraint alters the results from this model for the critical overdensity $\delta_\rmn{c}$ and the virial overdensity $\Delta_\rmn{v}$ more significantly than the other aforementioned effects found in numerical simulations.

In this article, I will show how the distribution of the gravitational shear's ellipticity and prolaticity at the position of density maxima can be derived from the statistics of Gaussian random fields by accounting for the \emph{two} aforementioned necessary constraints (vanishing first derivative and negative definite Hessian). In Section~\ref{sec:distributionEandP}, I give a short introduction to Gaussian random fields and show step-by-step how the targeted distribution can be calculated invoking the technique to derive number densities from multivariate Gaussians of the relevant quantities based on \citet{Bardeen1986} and how the constraint for the density's Hessian can be properly considered without diagonalisation. In Section~\ref{sec:results}, I compare the results from three distributions to each other. The first one incorporates only the constraint for the density's Hessian to be negative definite, the second one includes additionally the constraint for the vanishing first derivative of the density, and the third one is the distribution of \citet{Sheth2001} that does not take any of these constraints into account. Furthermore, I present the implications for the ellipsoidal-collapse model by \citet{Bond1996}, especially on the linear overdensity $\delta_\rmn{c}$ and the virial overdensity $\Delta_\rmn{v}$ in this section. Finally, I summarize the main results of this article in Section~\ref{sec:summary}.

\section{The conditional distribution for ellipticity and prolaticity}
\label{sec:distributionEandP}

In this section, I derive the conditional probability for the ellipticity and the prolaticity of the gravitational shear from the statistics of Gaussian random fields. First, I give a brief overview over the theoretical background of Gaussian random fields based on \citeauthor{Bardeen1986} (\citeyear{Bardeen1986}, see also \citealt{Angrick2009} for further information). Hence, I also adopt $F$ for the random field and $\bmath{\eta}=\nabla F$ and $\zeta_{ij}=\upartial_i \upartial_j F$ for its first and second derivatives, respectively.

\subsection{Theoretical background}
\label{subsec:TheoBackground}

An \emph{$n$-dimensional random field} $F$ assigns a set of random numbers to each point in $n$-dimensional space. A joint probability function can be declared for $m$ arbitrary points $\bmath{r}_j$ with $1\leq j\leq m$ as the probability that the field $F$, considered at the points $\bmath{r}_j$, has values between $F(\bmath{r}_j)$ and $F(\bmath{r}_j)+{\rm d}F(\bmath{r}_j)$.

A \emph{Gaussian random field} is a field whose joint probability functions are multivariate Gaussians. Let $y_i$ with $1\leq i\leq p$ be a set of Gaussian random variables with means $\langle y_i\rangle$ and $\Delta y_i\equiv y_i-\langle y_i\rangle$. The \emph{covariance matrix} $\tens{M}$ has the elements $M_{ij}\equiv\langle\Delta y_i\,\Delta y_j\rangle$, and the joint probability function of the Gaussian random variables is
\begin{equation}
\label{eq:multiGaussDef}
  P(y_1,\ldots,y_p)\,{\rm d}y_1\cdots{\rm d}y_p=
  \frac{1}{\sqrt{\left(2\pi\right)^p\det\tens{M}}}\,
  {\rm e}^{-Q}\,{\rm d}y_1\cdots{\rm d}y_p
\end{equation}
with the quadratic form
\begin{equation}
\label{eq:quadForm}
  Q\equiv\frac{1}{2}\sum_{i,j=1}^{p}\Delta y_i\left(\tens{M}^{-1}\right)_{ij}\Delta y_j.
\end{equation}
These random variables $y_i$ can, for example, not only include the field $F$, but also its derivatives, integrals or linear combinations thereof since they are also Gaussian.

In the following, I will focus on the \emph{density contrast} defined by $\delta(\bmath{r})\equiv[\rho(\bmath{r})-\rho_\rmn{b}]/\rho_\rmn{b}$ that is defined on the three-dimensional Euclidean space with coordinates $\bmath{r}=(x_1,x_2,x_3)^\rmn{T}$. Here, $\rho(\bmath{r})$ is the position-dependent density and $\rho_\rmn{b}$ is the cosmic background density. Hence, I will set $F\equiv\delta$ from now on. 

This homogeneous Gaussian random field with zero mean is fully characterized by its two-point correlation function $\xi(\bmath{r}_1,\bmath{r}_2)=\xi(|\bmath{r}_1-\bmath{r}_2|)\equiv\langle \delta(\bmath{r}_1)\,\delta(\bmath{r}_2)\rangle$ or equivalently its Fourier transform, the power spectrum $P(k)$. Its \emph{spectral moments} are defined by
\begin{equation}
\label{eq:specMoments}
\sigma^2_j\equiv\int_0^\infty\frac{\dd k}{2\pi^2}P(k)\,k^{2j+2}\,\hat{W}_R^2(k),
\end{equation}
where $\hat{W}_R(k)$ is the Fourier transform of the filter function $W_R(\bmath{r})$, and $R$ is its filter size. In the following, I adopt \emph{Gaussian filtering}, hence
\begin{equation}
\label{eq:filterFuncion}
W_R(\bmath{r})=\frac{1}{(2\pi)^{3/2}R^3}\exp\left(-\frac{|\bmath{r}|^2}{2R^2}\right)\quad\rmn{and}\quad\hat{W}(k)=\exp\left(-\frac{R^2 k^2}{2}\right).
\end{equation}
The zeroth-order spectral moment $\sigma_0^2$ is the power spectrum's \emph{variance}.

It is possible to scale the field $\delta$ as well as its derivatives with corresponding spectral moments so that they become dimensionless. I will denote these \emph{reduced fields} with a tilde, e.g.\ $\tilde{\delta}=\delta/\sigma_0$, $\tilde{\bmath{\eta}}=\bmath{\eta}/\sigma_1$, and $\tilde{\bmath{\zeta}}=\bmath{\zeta}/\sigma_2$, where now $\bmath{\eta}=\nabla\delta$ and $\zeta_{ij}=\upartial_i\upartial_j\delta$. A quantity that will be used more often in the next sections is a special combination of the first three spectral moments, $\gamma\equiv\sigma_1^2/(\sigma_0\,\sigma_2)$.

The elements $\varphi_{ij}$ of the gravitational-shear tensor $\bmath{\varphi}$ \citep{Zeldovich1970} contain the scaled second derivatives of the gravitational potential $\Phi$,
\begin{equation}
\label{eq:deformationTensor}
\varphi_{ij}=\frac{1}{4\pi G\rho_\rmn{b}}\frac{\upartial^2\Phi}{\upartial x_i\,\upartial x_j},
\end{equation}
where $G$ is Newton's constant, so that $\tr \bmath{\varphi}=\delta$ by Poisson's equation. I will denote the eigenvalues of $\bmath{\varphi}$ with $\lambda_i$, where $1\leq i\leq 3$. The ellipticity $e$ and the prolaticity $p$ are combinations of these eigenvalues,
\begin{equation}
\label{eq:defineEandP}
e\equiv\frac{\lambda_1-\lambda_3}{2\delta},\quad p\equiv\frac{\lambda_1-2\lambda_2+\lambda_3}{2\delta}\quad\rmn{with}\quad \delta=\lambda_1+\lambda_2+\lambda_3.
\end{equation}
Ordering the eigenvalues in the way $\lambda_1\geq\lambda_2\geq\lambda_3$, as it is commonly done in the literature, imposes the constraints $e\geq 0$ and $-e\leq p \leq e$ for $\delta\geq0$. Since a spherical configuration has $\lambda_1=\lambda_2=\lambda_3$, this is equivalent to $e=p=0$. While the ellipticity measures the anisotropy in the $x_1$-$x_3$-plane, the prolaticity quantifies the shape perpendicular to it: if $p<0$, the configuration is \emph{prolate}, for $p>0$ it is \emph{oblate}. Note that ellipticity and prolaticity are defined on the linear density field and should not be confused with the shape of the final halo.

Following \citet{Bardeen1986}, the conditional probability for the random variables $e$ and $p$ given a maximum of reduced height $\tilde{\delta}=\delta/\sigma_0$ can be expressed as
\begin{equation}
\label{eq:condProb}
P(e,p\,|\,\tilde{\delta},\rmn{max})=\frac{\int_{-\infty}^\infty\dd^6\tilde{\zeta}\,|\det\tilde{\bmath{\zeta}}|\,P(e,p,\tilde{\delta},\tilde{\bmath{\eta}}=\bmath{0},\tilde{\bmath{\zeta}})\,f(\bmath{\tilde{\zeta}})}{\int_{-\infty}^\infty\dd^6\tilde{\zeta}\,|\det\tilde{\bmath{\zeta}}|\,P(\tilde{\delta},\tilde{\bmath{\eta}}=\bmath{0},\tilde{\bmath{\zeta}})\,f(\bmath{\tilde{\zeta}})},
\end{equation}
where, as already mentioned above, $\tilde{\bmath{\eta}}$ is the reduced first derivative of $\delta$ and $\tilde{\bmath{\zeta}}$ is the corresponding reduced Hessian, which has only six independent components since it is symmetric. Note that only those subvolumes contribute to the integral for which $\tilde{\bmath{\zeta}}$ is negative definite, i.e.\ its three eigenvalues are negative. This is taken into account by the function $f(\bmath{\tilde{\zeta}})$ which has the property
\begin{equation}
\label{eq:fDefine}
f(\bmath{\tilde{\zeta}})=\begin{cases}
                           1\qquad\text{if }\bmath{\tilde{\zeta}}\text{ is negative definite}\\
                           0\qquad\text{else.}
                          \end{cases}
\end{equation}
I will discuss in Section~\ref{subsec:Sylvester} how the function $f(\bmath{\tilde{\zeta}})$ can be specified explicitly.

\subsection{The joint probability of relevant quantities}
\label{subsec:jointProb}

To calculate $P(e,p\,|\,\tilde{\delta},\rmn{max})$, I shall start with establishing the 15-dimensional joint probability of reduced variables $P(\tilde{\bmath{\varphi}},\tilde{\bmath{\eta}},\tilde{\bmath{\zeta}})\,\dd^6\tilde{\varphi}\,\dd^3\tilde{\eta}\,\dd^6\tilde{\zeta}$ to evaluate the numerator of equation~\eqref{eq:condProb}.

The correlation matrix $\tens{M}$ incorporates all auto- and cross-correlations of the various random variables. These are
\begin{align}
\label{eq:correlations}
\langle\tilde{\varphi}_{ij}\,\tilde{\varphi}_{kl}\rangle&=\frac{1}{15}\,\kappa_{ijkl},\quad & \langle\tilde{\eta}_i\,\tilde{\eta}_j\rangle&=\frac{1}{3}\,\delta_{ij},\\
\langle\tilde{\zeta}_{ij}\,\tilde{\zeta}_{kl}\rangle&=\frac{1}{15}\,\kappa_{ijkl},\quad & \langle\tilde{\varphi}_{ij}\,\tilde{\zeta}_{kl}\rangle&=-\frac{\gamma}{15}\,\kappa_{ijkl},
\end{align}
where $\delta_{ij}$ is the Kronecker symbol and the quantity $\kappa_{ijkl}$ is a combination of Kronecker symbols given by $\kappa_{ijkl}\equiv\delta_{ij}\delta_{kl}+\delta_{ik}\delta_{jl}+\delta_{il}\delta_{jk}$. All other correlations apart from those listed above vanish \citep[cf.\ also][]{Bardeen1986,Rossi2012}. Defining the vector of random variables
\begin{align}
\nonumber
&\bmath{y}=\\
\label{eq:randomVariables}
&(\tilde{\varphi}_{11},\tilde{\varphi}_{22},\tilde{\varphi}_{33},\tilde{\varphi}_{12},\tilde{\varphi}_{13},\tilde{\varphi}_{23},\tilde{\eta}_1,\tilde{\eta}_2,\tilde{\eta}_3,\tilde{\zeta}_{11},\tilde{\zeta}_{22},\tilde{\zeta}_{33},\tilde{\zeta}_{12},\tilde{\zeta}_{13},\tilde{\zeta}_{23})^\rmn{T},
\end{align}
the correlation matrix $\tens{M}$ looks like
\begin{equation}
\label{eq:M}
\tens{M}=\frac{1}{15}\begin{pmatrix}
   \tens{A} & \bmath{\emptyset} & \bmath{\emptyset} &-\gamma\cdot\tens{A} & \bmath{\emptyset} \\
\bmath{\emptyset} & \mathbfss{1} & \bmath{\emptyset} & \bmath{\emptyset} & -\gamma\cdot\mathbfss{1}\\
\bmath{\emptyset} & \bmath{\emptyset} & 5\cdot\mathbfss{1} & \bmath{\emptyset} & \bmath{\emptyset} \\
-\gamma\cdot\tens{A}& \bmath{\emptyset} & \bmath{\emptyset} & \tens{A} & \bmath{\emptyset} \\
\bmath{\emptyset} & -\gamma\cdot\mathbfss{1}& \bmath{\emptyset} & \bmath{\emptyset} & \mathbfss{1}
  \end{pmatrix},
\end{equation}
where the different submatrices are given by
\begin{equation}
\label{eq:submatricesM}
\tens{A}=\begin{pmatrix}
          3 & 1 & 1 \\
1 & 3 & 1 \\
1 & 1 & 3
         \end{pmatrix},\quad
\mathbfss{1}=\begin{pmatrix}
           1 & 0 & 0 \\
0 & 1 & 0 \\
0 & 0 & 1
          \end{pmatrix},\quad
\bmath{\emptyset}=\begin{pmatrix}
           0 & 0 & 0 \\
0 & 0 & 0 \\
0 & 0 & 0
\end{pmatrix}.
\end{equation}
The determinant of $\tens{M}$ is
\begin{equation}
\label{eq:determinant}
\det\tens{M}=\frac{2^4(1-\gamma^2)^6}{3^{15}\cdot5^{10}},
\end{equation}
and the inverse of $\tens{M}$ is
\begin{align}
\nonumber
&\tens{M}^{-1}=\\
\label{eq:Minv}
&\frac{3}{2(1-\gamma^2)}\begin{pmatrix}
               \tens{B} & \bmath{\emptyset} & \bmath{\emptyset} & \gamma\cdot\tens{B} & \bmath{\emptyset} \\
\bmath{\emptyset} & 10\cdot\mathbfss{1} & \bmath{\emptyset} &\bmath{\emptyset} & 10\gamma\cdot\mathbfss{1} \\
\bmath{\emptyset} & \bmath{\emptyset} & 2(1-\gamma^2)\cdot\mathbfss{1} & \bmath{\emptyset} & \bmath{\emptyset} \\
\gamma\cdot\tens{B} & \bmath{\emptyset} & \bmath{\emptyset} & \tens{B} & \bmath{\emptyset} \\
\bmath{\emptyset} & 10\gamma\cdot\mathbfss{1} & \bmath{\emptyset} & \bmath{\emptyset} & 10\cdot\mathbfss{1}
              \end{pmatrix},
\end{align}
where $\mathbfss{1}$ and $\bmath{\emptyset}$ are defined as above and
\begin{equation}
\label{eq:submatrixMInv}
\tens{B}=\begin{pmatrix}
          4 & -1 & -1 \\
-1 & 4 & -1 \\
-1 & -1 & 4
         \end{pmatrix}.
\end{equation}
Rotating the coordinate system such that $\tilde{\bmath{\varphi}}$ becomes diagonal and assuming that its eigenvalues are ordered like $\tilde{\lambda}_1\geq\tilde{\lambda}_2\geq\tilde{\lambda}_3$, \citet{Bardeen1986} showed that the volume element $\dd^6\tilde{\varphi}$ can be written as
\begin{equation}
\label{eq:volumeElement}
\dd^6\tilde{\varphi}=2\pi^2(\tilde{\lambda}_1-\tilde{\lambda}_2)(\tilde{\lambda}_1-\tilde{\lambda}_3)(\tilde{\lambda}_2-\tilde{\lambda}_3)\,\dd\tilde{\lambda}_1\,\dd\tilde{\lambda}_2\,\dd\tilde{\lambda}_3.
\end{equation}
Note that the two matrices $\tilde{\bmath{\varphi}}$ and $\tilde{\bmath{\zeta}}$ do not commute and hence cannot be diagonalized simultaneously.

Therefore, the off-diagonal elements of $\tilde{\bmath{\zeta}}$ do not vanish, whereas $\tilde{\varphi}_{12}=\tilde{\varphi}_{13}=\tilde{\varphi}_{23}=\tilde{\varphi}_{21}=\tilde{\varphi}_{31}=\tilde{\varphi}_{32}=0$, and the diagonal of $\tilde{\bmath{\varphi}}$ is populated by its eigenvalues $\tilde{\lambda}_i$ after the rotation.

The circumstance that $\bmath{\varphi}$ and $\bmath{\zeta}$ do not commute can be shown as follows. The elements of the density's Hessian $\zeta_{ij}$ are given by
\begin{equation}
\label{eq:defineHessian}
\zeta_{ij}=\frac{\upartial^2\delta}{\upartial x_i\,\upartial x_j}=\frac{\upartial^2}{\upartial x_i\,\upartial x_j}\sum_{k=1}^3\varphi_{kk}\equiv\sum_{k=1}^3\varphi_{ijkk},
\end{equation}
where $\varphi_{ijkl}$ is chosen as an abbreviation for the scaled fourth derivative of the gravitational potential $\Phi$ analogous to equation~\eqref{eq:deformationTensor}. The commutator of the two matrices $\bmath{\varphi}$ and $\bmath{\zeta}$ is defined as $[\bmath{\varphi},\bmath{\zeta}]\equiv\bmath{\varphi\zeta}-\bmath{\zeta\varphi}$. Written in components, it is given by
\begin{equation}
\label{eq:commutator}
\begin{split}
(\bmath{\varphi\zeta}-\bmath{\zeta\varphi})_{ij}&=\sum_{k=1}^3\varphi_{ik}\sum_{l=1}^3\varphi_{kjll}-\sum_{k=1}^3\sum_{l=1}^3\varphi_{ikll}\,\varphi_{kj}\\
&=\sum_{k,l=1}^3(\varphi_{ik}\,\varphi_{kjll}-\varphi_{jk}\,\varphi_{kill})
\begin{cases}
=0 &\quad\rmn{if}~i=j,\\
\neq 0 &\quad\rmn{else,}
\end{cases}
\end{split}
\end{equation}
where I have used that $\varphi_{ij}=\varphi_{ji}$ and $\varphi_{ijkk}=\varphi_{jikk}$. Hence, the two matrices $\bmath{\varphi}$ and $\bmath{\zeta}$ do not commute.

Switching from the eigenvalues $\tilde{\lambda}_i$ to the new variables $\tilde{\delta}$, $e$ and $p$ (cf.\ equation~\ref{eq:defineEandP}), the Jacobi determinant of the transformation introduces an additional factor so that
\begin{equation}
\label{eq:JacobiDeterminant}
\dd\tilde{\lambda}_1\,\dd\tilde{\lambda}_2\,\dd\tilde{\lambda}_3=\frac{2}{3}\tilde{\delta}^2\,\dd\tilde{\delta}\,\dd e\,\dd p.
\end{equation}

Starting from equations~\eqref{eq:multiGaussDef} and \eqref{eq:quadForm} with the random variables \eqref{eq:randomVariables}, inverse \eqref{eq:Minv} of the correlation matrix and the correlation matrix' determinant \eqref{eq:determinant}, one can construct the multivariate Gaussian probability distribution \eqref{eq:multiGaussDef}. Then, I rotate the coordinate system such that the matrix $\tilde{\bmath{\varphi}}$ becomes diagonal and switch from the three eigenvalues $\tilde{\lambda}_i$ to the new variables $\tilde{\delta}$, $e$, and $p$, taking into account that the differentials transform like shown in equations~\eqref{eq:volumeElement} and \eqref{eq:JacobiDeterminant}. Finally, since I concentrate on maxima, I set $\bmath{\eta}=\bmath{0}$ so that
\begin{equation}
\label{eq:detEta}
\dd^3\eta=|\det\bmath{\zeta}|\,\dd^3 r\quad\rmn{and}\quad\dd^3\tilde{\eta}=\frac{\sigma_2^3}{\sigma_1^3}\,|\det\tilde{\bmath{\zeta}}|\,\dd^3r.
\end{equation}

In that way, the integrand in the numerator of equation~\eqref{eq:condProb} can be written as
\begin{multline}
\label{eq:jointProb}
P(e,p,\tilde{\delta},\tilde{\bmath{\eta}}=\bmath{0},\tilde{\bmath{\zeta}})\,|\det\tilde{\bmath{\zeta}}|=\\
\left(\frac{3}{2}\right)^{13/2}\frac{5^5}{\pi^{11/2}(1-\gamma^2)^3}\,e\,(e^2-p^2)\,\tilde{\delta}^5\exp\,\Biggl\{\frac{1}{2(\gamma^2-1)}\left[3(2\tilde{\delta}_\zeta^2-5\tilde{\Delta})\right.\\
+\left.2\gamma\tilde{\delta}\tilde{\delta}_\zeta(1+15ee_\zeta+5pp_\zeta)+\tilde{\delta}^2(1+15e^2+5p^2)\right]\Biggr\}\,|\det\tilde{\bmath{\zeta}}|,
\end{multline}
where I have introduced the abbreviations
\begin{equation}
\label{eq:abbreviations}
\begin{split}
\tilde{\Delta}&\equiv\tilde{\zeta}_{11}\tilde{\zeta}_{22}+\tilde{\zeta}_{11}\tilde{\zeta}_{33}+\tilde{\zeta}_{22}\tilde{\zeta}_{33}-\tilde{\zeta}_{12}^2-\tilde{\zeta}_{13}^2-\tilde{\zeta}_{23}^2,\\
\tilde{\delta}_\zeta&\equiv\tilde{\zeta}_{11}+\tilde{\zeta}_{22}+\tilde{\zeta}_{33},\quad e_\zeta\equiv\frac{\tilde{\zeta}_{11}-\tilde{\zeta}_{33}}{2\tilde{\delta}_\zeta}, \quad p_\zeta\equiv\frac{\tilde{\zeta}_{11}-2\tilde{\zeta}_{22}+\tilde{\zeta}_{33}}{2\tilde{\delta}_\zeta},
\end{split}
\end{equation}
and with
\begin{equation}
\label{eq:detZeta}
\det\tilde{\bmath{\zeta}}=\tilde{\zeta}_{11}\tilde{\zeta}_{22}\tilde{\zeta}_{33}+2\tilde{\zeta}_{12}\tilde{\zeta}_{13}\tilde{\zeta}_{23}-\tilde{\zeta}_{11}\tilde{\zeta}_{23}^2-\tilde{\zeta}_{22}\tilde{\zeta}_{13}^2-\tilde{\zeta}_{33}\tilde{\zeta}_{12}^2.
\end{equation}

\subsection{Sylvester's criterion}
\label{subsec:Sylvester}

To evaluate the numerator of equation~\eqref{eq:condProb}, I have to integrate over those parts of the volume spanned by the matrix elements of $\tilde{\bmath{\zeta}}$ for which the latter is negative definite. This can be achieved by taking into account the following criterion. A Hermitian matrix is positive definite if and only if all its leading principal minors are positive, where the $i$th leading principal minor is the determinant of its upper left $i\times i$ submatrix. This is known as \emph{Sylvester's criterion}. Since the negative of a positive definite and Hermitian matrix is negative definite, this implies that all odd principal minors have to be negative and all even principal minors have to be positive. 

Thus, the function $f(\bmath{\tilde{\zeta}})$ introduced in equation~\eqref{eq:condProb} can be specified as
\begin{equation}
\label{eq:f}
f(\tilde{\bmath{\zeta}})=\begin{cases}
                   1 & \rmn{if}~~\tilde{\zeta}_{11}<0\,\wedge\, \begin{vmatrix}
                \tilde{\zeta}_{11} & \tilde{\zeta}_{12} \\
\tilde{\zeta}_{12} & \tilde{\zeta}_{22}
               \end{vmatrix}>0\, \wedge\, \begin{vmatrix}
\tilde{\zeta}_{11} & \tilde{\zeta}_{12} & \tilde{\zeta}_{13} \\
\tilde{\zeta}_{12} & \tilde{\zeta}_{22} & \tilde{\zeta}_{23} \\
\tilde{\zeta}_{13} & \tilde{\zeta}_{23} & \tilde{\zeta}_{33} 
\end{vmatrix}<0 \\
0 & \rmn{else.}
\end{cases}
\end{equation}

When integrating equation~\eqref{eq:jointProb} numerically over the various matrix elements, I also multiply it by the function $f(\tilde{\bmath{\zeta}})$
and integrate over all six independent elements from $-\infty$ to $\infty$.

\subsection{The number density of maxima in the density field}
\label{subsec:num}
In analogy to the way I have derived equation~\eqref{eq:jointProb}, it is possible to derive the joint probability $P(\tilde{\delta},\tilde{\bmath{\eta}},\tilde{\bmath{\zeta}})\,\dd\tilde{\delta}\,\dd^3\tilde{\eta}\,\dd^6\tilde{\zeta}$. Apart from the correlations $\langle \tilde{\eta}_i\,\tilde{\eta}_j\rangle$ and $\langle \tilde{\zeta}_{ij}\,\tilde{\zeta}_{kl}\rangle$, which are already given in equation~\eqref{eq:correlations}, the auto-correlation of the density contrast, $\langle\tilde{\delta}\,\tilde{\delta}\rangle=1$ and the cross-correlation $\langle\tilde{\delta}\,\tilde{\zeta}_{ij}\rangle=-\gamma\,\delta_{ij}/3$ are the only non-vanishing contributions that are needed for the derivation. The vector of random variables that enters equation~\eqref{eq:multiGaussDef} in this case has 10 components and can be written by
\begin{equation}
\label{eq:vectorRand2}
\bmath{y}=(\tilde{\delta},\tilde{\eta}_1,\tilde{\eta}_2,\tilde{\eta}_3,\tilde{\zeta}_{11},\tilde{\zeta}_{22},\tilde{\zeta}_{33},\tilde{\zeta}_{12},\tilde{\zeta}_{13},\tilde{\zeta}_{23})^\rmn{T},
\end{equation}
so that the correlation matrix looks like
\begin{equation}
\label{eq:M2}
\tens{M}=\frac{1}{15}\begin{pmatrix}
1 & \bmath{O}^\rmn{T} & -5\gamma\cdot\bmath{I}^\rmn{T} & \bmath{O}^\rmn{T} \\
\bmath{O} & 5\cdot\mathbfss{1} & \bmath{\emptyset} & \bmath{\emptyset} \\
-5\gamma\cdot\bmath{I} & \bmath{\emptyset} & \tens{A} & \bmath{\emptyset} \\
\bmath{O} & \bmath{\emptyset} & \bmath{\emptyset} & 3\cdot\mathbfss{1}
  \end{pmatrix},
\end{equation}
where $\bmath{\emptyset}$, $\mathbfss{1}$, and $\tens{A}$ are defined in equation~\eqref{eq:submatricesM}, whereas $\bmath{O}$ and $\bmath{I}$ are vectors of the form
\begin{equation}
\label{eq:subVector2}
\bmath{O}=\begin{pmatrix}
                0 \\
0 \\
0
               \end{pmatrix}\quad\rmn{and}\quad\bmath{I}=\begin{pmatrix}
1 \\
1 \\
1
\end{pmatrix},
\end{equation}
respectively. Inverting $\tens{M}$ leads to
\begin{equation}
\label{eq:Minv2}
\tens{M}^{-1}=\frac{1}{2(1-\gamma^2)}\begin{pmatrix}
2 & \bmath{O}^\rmn{T} & 2\gamma\cdot\bmath{I}^\rmn{T} & \bmath{O}^\rmn{T} \\
\bmath{O} & 6\cdot\mathbfss{1} & \bmath{\emptyset} & \bmath{\emptyset} \\
2\gamma\cdot\bmath{I} & \bmath{\emptyset} & \tens{C} & \bmath{\emptyset} \\
\bmath{O} & \bmath{\emptyset} & \bmath{\emptyset} & 30\cdot\mathbfss{1}
                                  \end{pmatrix},
\end{equation}
where
\begin{equation}
\label{eq:tensC}
\tens{C}=\begin{pmatrix}
12-10\gamma^2 & 5\gamma^2-3 & 5\gamma^2-3 \\
5\gamma^2-3 & 12-10\gamma^2 & 5\gamma^2-3 \\
5\gamma^2-3 & 5\gamma^2-3 & 12-10\gamma^2
         \end{pmatrix}.
\end{equation}

At this point, it is most efficient to rotate the coordinate system into the eigensystem of $\tilde{\bmath{\zeta}}$. Note that this rotation is different from the one applied when calculating equation~\eqref{eq:jointProb}, since there I aimed at diagonalising $\tilde{\bmath{\varphi}}$ instead of $\tilde{\bmath{\zeta}}$ to introduce the ellipticity $e$ and the prolaticity $p$, which are defined in the eigensystem of $\tilde{\bmath{\varphi}}$. The calculation in this section, however, corresponds to deriving the number density of maxima of a certain height $\tilde{\delta}$ (see below), which is a quantity independent of the choice of the coordinate system's rotation so that I am free to choose a different rotation from the one applied to derive equation~\eqref{eq:jointProb}.

Using the ordered eigenvalues $\tilde{\lambda}_{\zeta,1}\geq\tilde{\lambda}_{\zeta,2}\geq\tilde{\lambda}_{\zeta,3}$ of the density's reduced Hessian $\tilde{\bmath{\zeta}}$, the volume element $\dd^6\tilde{\zeta}$ can be written after the diagonalisation analogously to equation~\eqref{eq:volumeElement} as
\begin{equation}
\label{eq:transformZeta}
\dd^6\tilde{\zeta}=2\pi^2(\tilde{\lambda}_{\zeta,1}-\tilde{\lambda}_{\zeta,2})(\tilde{\lambda}_{\zeta,1}-\tilde{\lambda}_{\zeta,3})(\tilde{\lambda}_{\zeta,2}-\tilde{\lambda}_{\zeta,3})\,\dd\tilde{\lambda}_{\zeta,1}\,\dd\tilde{\lambda}_{\zeta,2}\,\dd\tilde{\lambda}_{\zeta,3}.
\end{equation}

Switching to the new variables $e_\zeta$, $p_\zeta$ and $\tilde{\delta}$, introduced in equation~\eqref{eq:abbreviations}, which are functions of the eigenvalues $\tilde{\lambda}_{\zeta,i}$ after the rotation,
\begin{equation}
\label{eq:newVariablesZeta}
\begin{split}
e_\zeta&=\frac{\tilde{\lambda}_{\zeta,1}-\tilde{\lambda}_{\zeta,3}}{2\tilde{\delta}_\zeta},\quad p_\zeta=\frac{\tilde{\lambda}_{\zeta,1}-2\tilde{\lambda}_{\zeta,2}+\tilde{\lambda}_{\zeta,3}}{2\tilde{\delta}_\zeta}\quad\rmn{with}\\
\tilde{\delta}_\zeta&=\tilde{\lambda}_{\zeta,1}+\tilde{\lambda}_{\zeta,2}+\tilde{\lambda}_{\zeta,3},
\end{split}
\end{equation}
introduces an additional factor through the Jacobian of the transformation,
\begin{equation}
\label{eq:JacobianZeta}
\dd\tilde{\lambda}_{\zeta,1}\,\dd\tilde{\lambda}_{\zeta,2}\,\dd\tilde{\lambda}_{\zeta,3}=\frac{2}{3}\tilde{\delta}^2_\zeta\,\dd\tilde{\delta}_\zeta\,\dd e_\zeta\,\dd p_\zeta.
\end{equation}
The last step again is to set $\bmath{\eta}=\bmath{0}$ so that
\begin{equation}
\label{eq:detAfterRotation}
\begin{split}
\dd^3\tilde{\eta}&=\frac{\sigma_2^3}{\sigma_1^3}\,|\det\tilde{\bmath{\zeta}}|\,\dd^3r=-\frac{\sigma_2^3}{\sigma_1^3}\,\lambda_{\zeta,1}\,\lambda_{\zeta,2}\,\lambda_{\zeta,3}\,\dd^3r \\
&=\left(\frac{\tilde{\delta}_\zeta\sigma_2}{3\sigma_1}\right)^3[9e_\zeta^2-(1+p_\zeta)^2](1-2p_\zeta)\,\dd^3 r,
\end{split}
\end{equation}
where I have used that $|\det\tilde{\bmath{\zeta}}|$ is negative if $\tilde{\bmath{\zeta}}$ is negative definite.

Applying all the various steps mentioned above, the integrand in the denominator of equation~\eqref{eq:condProb} can be written as
\begin{multline}
\label{eq:numerator}
P(\tilde{\delta},\tilde{\bmath{\eta}}=\bmath{0},\tilde{\bmath{\zeta}})\,|\det\tilde{\bmath{\zeta}}|=\\
\frac{25\sqrt{15}}{8\pi^3\sqrt{1-\gamma^2}}e_\zeta(e_\zeta^2-p_\zeta^2)[9e_\zeta^2-(1+p_\zeta)^2](1-2p_\zeta)\tilde{\delta}_\zeta^8\\
\times\exp\left[\frac{(\tilde{\delta}^2+\tilde{\delta}_\zeta^2+2\gamma\tilde{\delta}\tilde{\delta}_\zeta+(1-\gamma^2)(15e_\zeta^2+5p_\zeta)\tilde{\delta}_\zeta^2}{2(\gamma^2-1)}\right].
\end{multline}
To evaluate the denominator of equation~\eqref{eq:condProb}, I have to integrate over the eigenvalues $\tilde{\lambda}_{\zeta,i}$ from $-\infty$ to 0, or, since I have transformed to the new variables $\tilde{\delta}_\zeta$, $e_\zeta$, and $p_\zeta$, integrate over them in the proper range. The condition $0\geq\tilde{\lambda}_{\zeta,1}\geq\tilde{\lambda}_{\zeta,2}\geq\tilde{\lambda}_{\zeta,3}$ translated to the new variables reads
\begin{multline}
\label{eq:conditions}
\tilde{\delta}_\zeta\leq 0 \wedge\left[\left(-\frac{1}{2}\leq e_\zeta\leq-\frac{1}{4}\wedge-1-3e_\zeta\leq p_\zeta\leq -e_\zeta\right)\right.\\
\vee\left.\left(-\frac{1}{4}<e_\zeta\leq 0 \wedge e_\zeta\leq p_\zeta\leq-e_\zeta\right)\right].
\end{multline}
Note that the variables $\tilde{\delta}_\zeta$ and $e_\zeta$ are negative in contrast to the analogous variables $\tilde{\delta}$ and $e$ for the gravitational-shear tensor.

It turns out that the integrations over $e_\zeta$ and $p_\zeta$ can be done analytically, while the integration over $\tilde{\delta}_\zeta$ has to be done numerically. The denominator of equation~\eqref{eq:condProb} is equal to the number density of maxima $n_\rmn{max}(\tilde{\delta})$ of the reduced density field $\tilde{\delta}$ if multiplied by a factor $\sigma_2^3/\sigma_1^3$. The latter can be written as
\begin{equation}
\label{eq:numDensMax}
n_\rmn{max}(\tilde{\delta})=\frac{\sigma_2^3}{1200\pi^3\sigma_1^3\sqrt{3(1-\gamma^2)}}\int_{-\infty}^0\dd\tilde{\delta}_\zeta\,u\left(\sqrt{5}v+25 \sqrt{2\pi}\,w\right)
\end{equation}
with
\begin{equation}
\label{eq:numDensSubs}
\begin{split}
u&\equiv\exp\left[\frac{\tilde{\delta}^2+2\gamma\,\tilde{\delta}\,\tilde{\delta}_\zeta+\tilde{\delta}_\zeta^2(6-5\gamma^2)}{2(\gamma^2-1)}\right],\\
v&\equiv10\tilde{\delta}_\zeta^2-32+(155\tilde{\delta}_\zeta^2+32)\exp\left(\frac{15\tilde{\delta}_\zeta^2}{8}\right),\\
w&\equiv\tilde{\delta}_\zeta\,(\tilde{\delta}_\zeta^2-3)\exp\left(\frac{5\tilde{\delta}_\zeta^2}{2}\right)\left[\erf\left(\frac{1}{2}\sqrt{\frac{5}{2}}\tilde{\delta}_\zeta\right)+\erf\left(\sqrt{\frac{5}{2}}\tilde{\delta}_\zeta\right)\right].
\end{split}
\end{equation}

Now all the ingredients to express the probability to find the ellipticity $e$ and the prolaticity $p$ of the gravitational-shear field in the range $[e,e+\dd e]$ and $[p,p+\dd p]$ given a density peak of height $\delta$ are at hand. To summarize, after scaling it to the reduced density $\tilde{\delta}=\delta/\sigma_0$, the probability is given by the ratio \eqref{eq:condProb}, where the numerator is expressed by equation~\eqref{eq:jointProb} together with equations~\eqref{eq:abbreviations} and \eqref{eq:detZeta}, and the denominator by equations~\eqref{eq:numDensMax} and \eqref{eq:numDensSubs} except the factor $\sigma_2^3/\sigma_1^3$. The function $f(\tilde{\bmath{\zeta}})$ defined in equation~\eqref{eq:f} offers a nice way to take into account the constraint that equation~\eqref{eq:jointProb} has to be integrated only over those subvolumes for which the matrix $\tilde{\bmath{\zeta}}$ is negative definite.

\subsection{Comparison to the ansatz of Rossi}
\label{subsec:Rossi}

\citet{Rossi2012} derives a conditional probability for the reduced shear tensor $\tilde{\bmath{\varphi}}$ given a positive definite Hessian $\tilde{\bmath{\zeta}}$ (see his equations~18 and 19). In the language of this paper, his probability looks like
\begin{equation}
\label{eq:Rossi}
P(\tilde{\bmath{\varphi}}\,|\,\tilde{\bmath{\zeta}}\text{ pos.\ def.})=\frac{\int_{-\infty}^\infty\dd^6\tilde{\zeta}\,P(\tilde{\bmath{\zeta}})\,P(\tilde{\bmath{\varphi}}\,|\,\tilde{\bmath{\zeta}})\,g(\bmath{\tilde{\zeta}})}{\int_{-\infty}^\infty\dd^6\tilde{\zeta}\,P(\tilde{\bmath{\zeta}})\,g(\bmath{\tilde{\zeta}})},
\end{equation}
where $P(\tilde{\bmath{\varphi}}\,|\,\tilde{\bmath{\zeta}})=P(\tilde{\bmath{\varphi}},\tilde{\bmath{\zeta}})/P(\tilde{\bmath{\zeta}})$, and $g(\bmath{\tilde{\zeta}})$ accounts for integration over the subvolume for which $\tilde{\bmath{\zeta}}$ is \emph{positive} definite and can thus be similarly declared as $f(\bmath{\tilde{\zeta}})$ in equation~\eqref{eq:f} with the difference that \emph{all} submatrices need to have positive determinants. Both $P(\tilde{\bmath{\varphi}}\,|\,\tilde{\bmath{\zeta}})$ and $P(\tilde{\bmath{\zeta}})$ can be written as an expression analogous to the distribution of the gravitational-shear tensor's eigenvalues derived by \citet{Doroshkevich1970}.

However, it is not possible to diagonalize both matrices $\tilde{\bmath{\varphi}}$ and $\tilde{\bmath{\zeta}}$ simultaneously since the commutator of the two does not vanish. Hence, the aim of \citet{Rossi2013} is that this ansatz can be used to determine the probability $P(e,p\,|\,\delta,\rmn{max})$ from equation~\eqref{eq:Rossi}, implying that the eigenvalues of both the former matrices can be used. This, however, is not possible since only one of them can be diagonalized at once, it is not obvious how the six-dimensional integration over \emph{all} entries of the matrix $\tilde{\bmath{\zeta}}$ can be circumvented.

Furthermore, equation~\eqref{eq:Rossi} seems incomplete in the sense that it neglects the constraint that the first derivative $\bmath{\eta}$ has to vanish for a maximum. \citet{Rossi2012,Rossi2013} only takes into account the Hessian which has to be positive definite,\footnote{Actually, it should be \emph{negative} definite for a maximum. However, a homogeneous Gaussian random field with zero mean is symmetric in the sense that integration over subvolumes which are either positive or negative definite yields the same result.} and therefore, his probability is not only based on maxima in the density field but also on areas for which the first derivative is non-zero. Equation~\eqref{eq:condProb} is more specific in that respect since it takes the constraint $\bmath{\eta}=\bmath{0}$ explicitly into account \citep[see also][]{Bardeen1986}.

To quantify the difference between accounting for this additional constraint or not, I will also evaluate the conditional probability that only incorporates the constraint on the density's Hessian. In this case, equation~\eqref{eq:condProb} simplifies to
\begin{equation}
\label{eq:condProbWithout}
P(e,p\,|\,\tilde{\delta},\bmath{\tilde{\zeta}}\text{ neg.\ def.})=\frac{\int_{-\infty}^\infty\dd^6\tilde{\zeta}\,P(e,p,\tilde{\delta},\tilde{\bmath{\zeta}})\,f(\bmath{\tilde{\zeta}})}{\int_{-\infty}^\infty\dd^6\tilde{\zeta}\,P(\tilde{\delta},\tilde{\bmath{\zeta}})\,f(\bmath{\tilde{\zeta}})},
\end{equation}
where I integrate over the subvolumes for which $\tilde{\bmath{\zeta}}$ is \emph{negative} definite for consistency. Note the absence of the factor $|\det\tilde{\bmath{\zeta}}|$ in both integrals in equation~\eqref{eq:condProbWithout} in comparison to equation~\eqref{eq:condProb} which originated from the property that $\eta_i=\sum_j \zeta_{ij}\,x_j$ in the infinitesimal volume around a maximum and therefore $\dd^3\eta=|\det\bmath{\zeta}|\,\dd^3r$.

Since $P(e,p\,|\,\tilde{\delta},\bmath{\tilde{\zeta}}\text{ neg.\ def.})$ can be derived analogously to $P(e,p\,|\,\tilde{\delta},\rmn{max.})$ except that the first derivative $\bmath{\eta}$ is missing since no constraints are applied to it, I will only present the final result. The integrand in the numerator of equation~\eqref{eq:condProbWithout} is given by
\begin{multline}
\label{eq:numeratorWithout}
P(e,p,\tilde{\delta},\tilde{\bmath{\zeta}})=\\
\frac{15^5}{2^5\pi^4(1-\gamma^2)^3}\,e\,(e^2-p^2)\,\tilde{\delta}^5\exp\,\Biggl\{\frac{1}{2(\gamma^2-1)}\left[3(2\tilde{\delta}_\zeta^2-5\tilde{\Delta})\right.\\
+\left.2\gamma\tilde{\delta}\tilde{\delta}_\zeta(1+15ee_\zeta+5pp_\zeta)+\tilde{\delta}^2(1+15e^2+5p^2)\right]\Biggr\},
\end{multline}
while the integrand in the denominator can be written as
\begin{multline}
\label{eq:denominatorWithout}
P(\tilde{\delta},\tilde{\bmath{\zeta}})=\frac{225\sqrt{5}}{2\pi^{3/2}\sqrt{2(1-\gamma^2)}}e_\zeta(e_\zeta^2-p_\zeta^2)\tilde{\delta}_\zeta^5\\
\times\exp\left[\frac{(\tilde{\delta}^2+\tilde{\delta}_\zeta^2+2\gamma\tilde{\delta}\tilde{\delta}_\zeta+(1-\gamma^2)(15e_\zeta^2+5p_\zeta)\tilde{\delta}_\zeta^2}{2(\gamma^2-1)}\right].
\end{multline}
Equations~\eqref{eq:numeratorWithout} and \eqref{eq:denominatorWithout} deviate from equations~\eqref{eq:jointProb} and \eqref{eq:numerator} only in two ways. (a) The normalizations differ because the vanishing first derivative was not taken into account in equations~\eqref{eq:numeratorWithout} and \eqref{eq:denominatorWithout}. This difference, however, drops out when taking their ratio. (b) The factor $|\det\tilde{\bmath{\zeta}}|$ is not present in both equations. In contrast to the previous point, this additional factor does not vanish when taking the ratio since in both the numerator and the denominator, an integration over the reduced Hessian $\tilde{\bmath{\zeta}}$ still has to be carried out.

Integrating equation~\eqref{eq:denominatorWithout} over $e_\zeta$ and $p_\zeta$ in the range \eqref{eq:conditions} leads to
\begin{equation}
\label{eq:downIntegrate}
\int_{-\infty}^\infty\dd^6\tilde{\zeta}\,P(\tilde{\delta},\tilde{\bmath{\zeta}})\,f(\bmath{\tilde{\zeta}})=\frac{1}{8\pi^{3/2}\sqrt{1-\gamma^2}}\int_{-\infty}^0\dd\tilde{\delta}_\zeta\,b(3\sqrt{10}c-2\sqrt{\pi} d)
\end{equation}
with
\begin{equation}
\label{eq:downIntegrateSubs}
\begin{split}
b&\equiv\exp\left[\frac{\tilde{\delta}(\tilde{\delta}+2\gamma\tilde{\delta}_\zeta)}{2(\gamma^2-1)}\right],\\
c&\equiv\tilde{\delta}_\zeta\exp\left[\frac{(9-5\gamma^2)\tilde{\delta}_\zeta^2}{8(\gamma^2-1)}\right],\\
d&\equiv\exp\left[\frac{\tilde{\delta}_\zeta^2}{2(\gamma^2-1)}\right]\left[\erf\left(\frac{1}{2}\sqrt{\frac{5}{2}}\tilde{\delta}_\zeta\right)+\erf\left(\sqrt{\frac{5}{2}}\tilde{\delta}_\zeta\right)\right],
\end{split}
\end{equation}
which differs clearly from equations~\eqref{eq:numDensMax} and \eqref{eq:numDensSubs}.

\section{Results}
\label{sec:results}

\begin{figure*}
\centering
\includegraphics[width=0.495\textwidth]{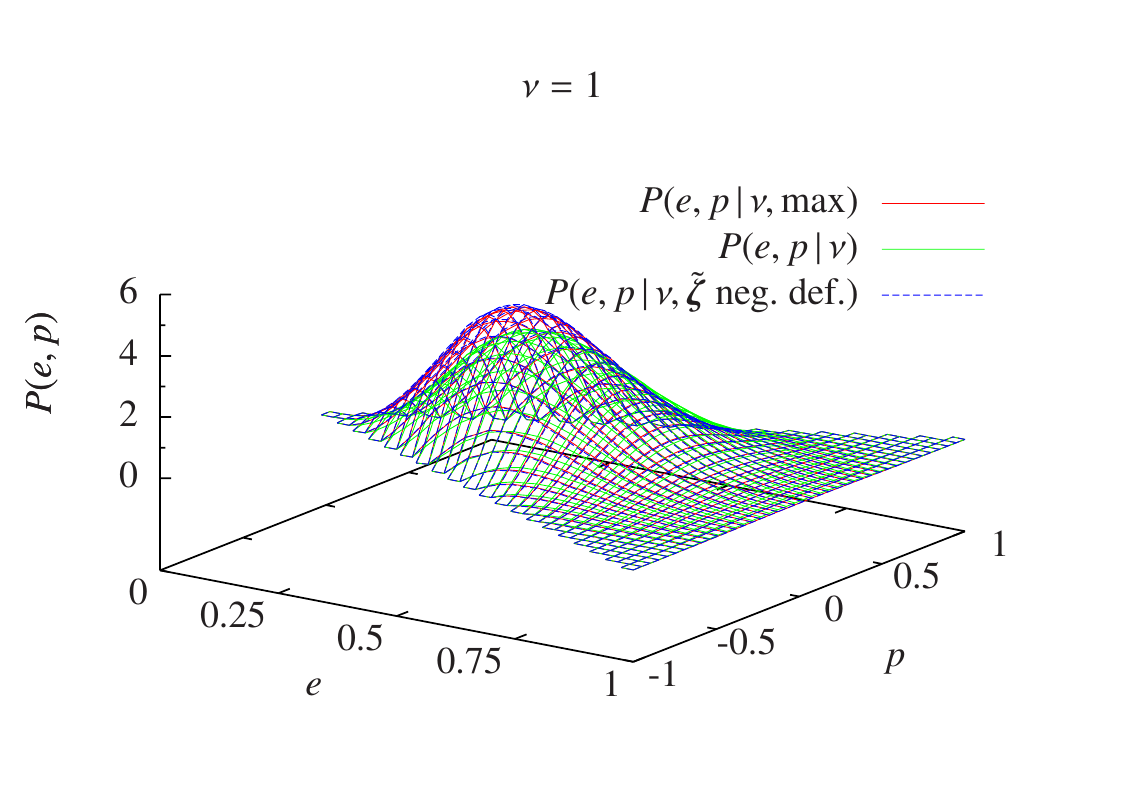}~\includegraphics[width=0.495\textwidth]{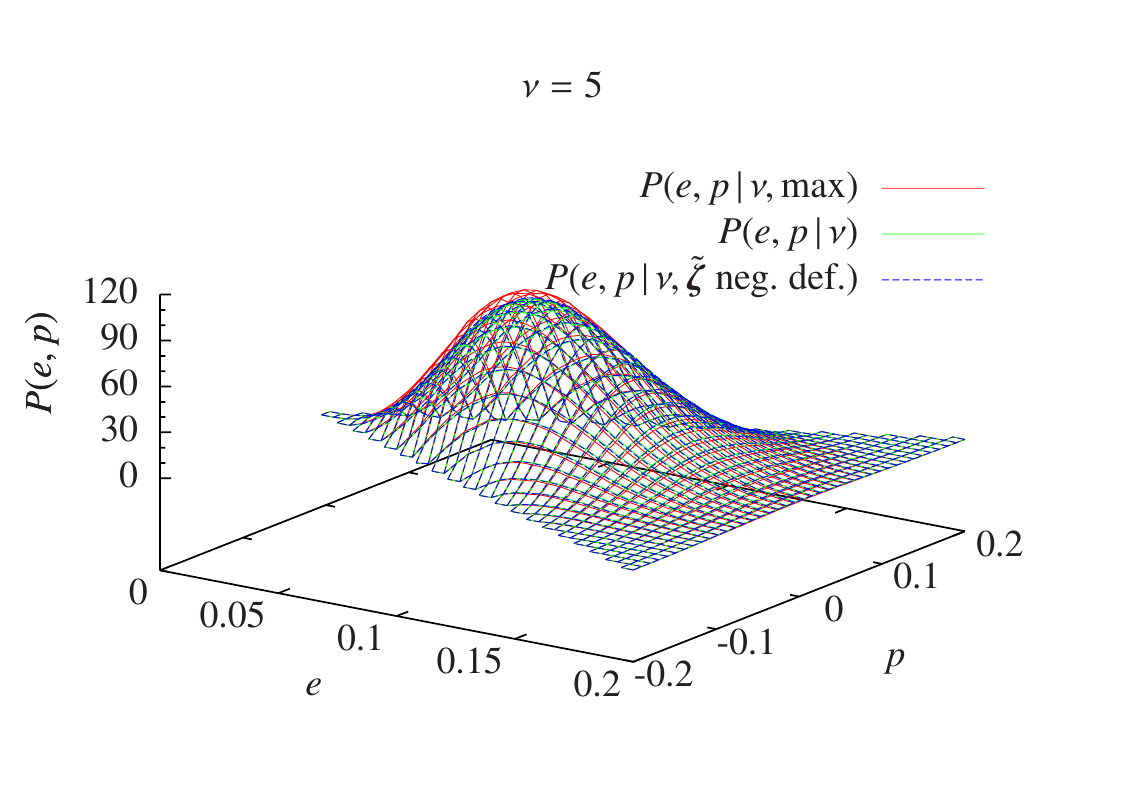}\\
\includegraphics[width=0.495\textwidth]{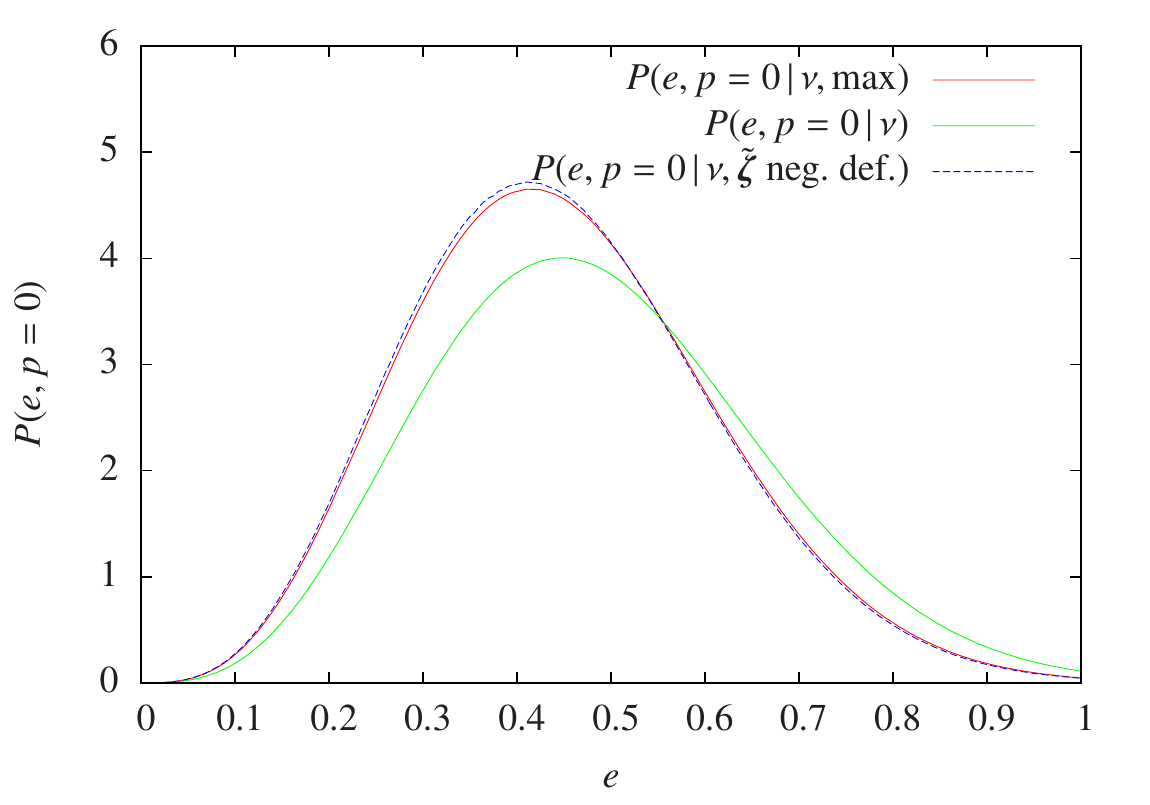}~\includegraphics[width=0.495\textwidth]{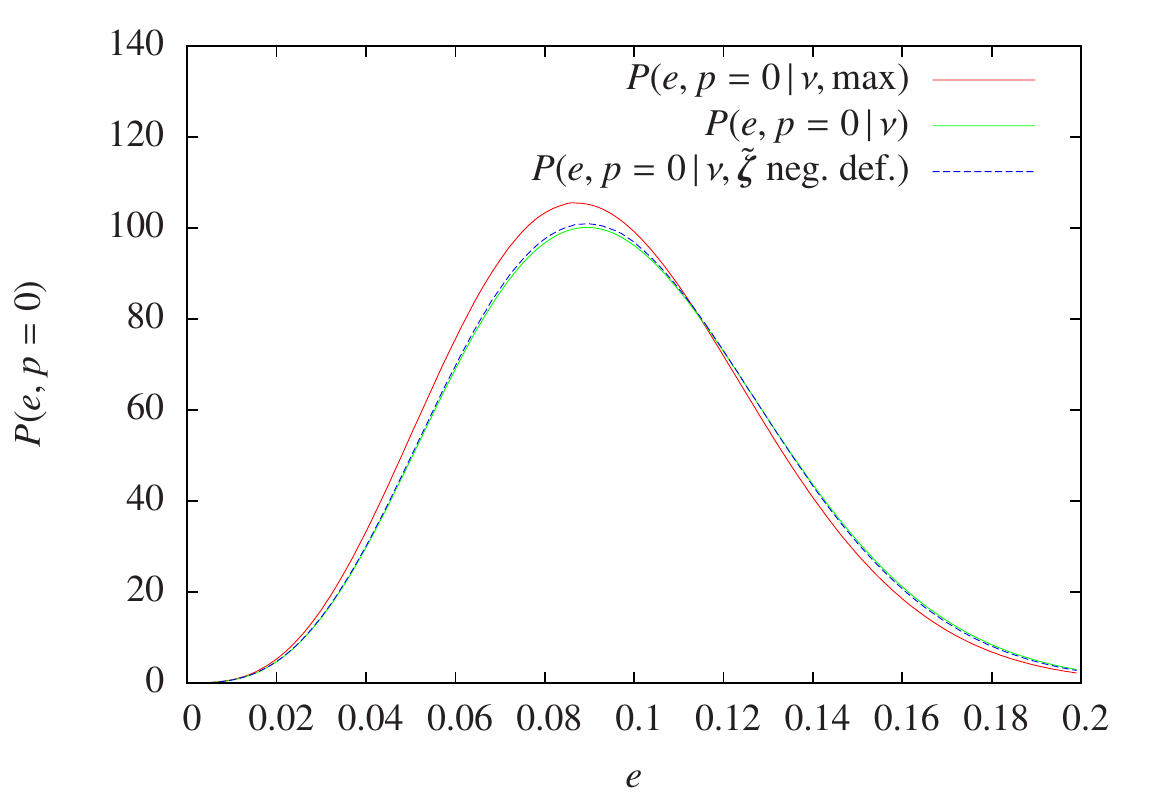}\\[2mm]
\includegraphics[width=0.495\textwidth]{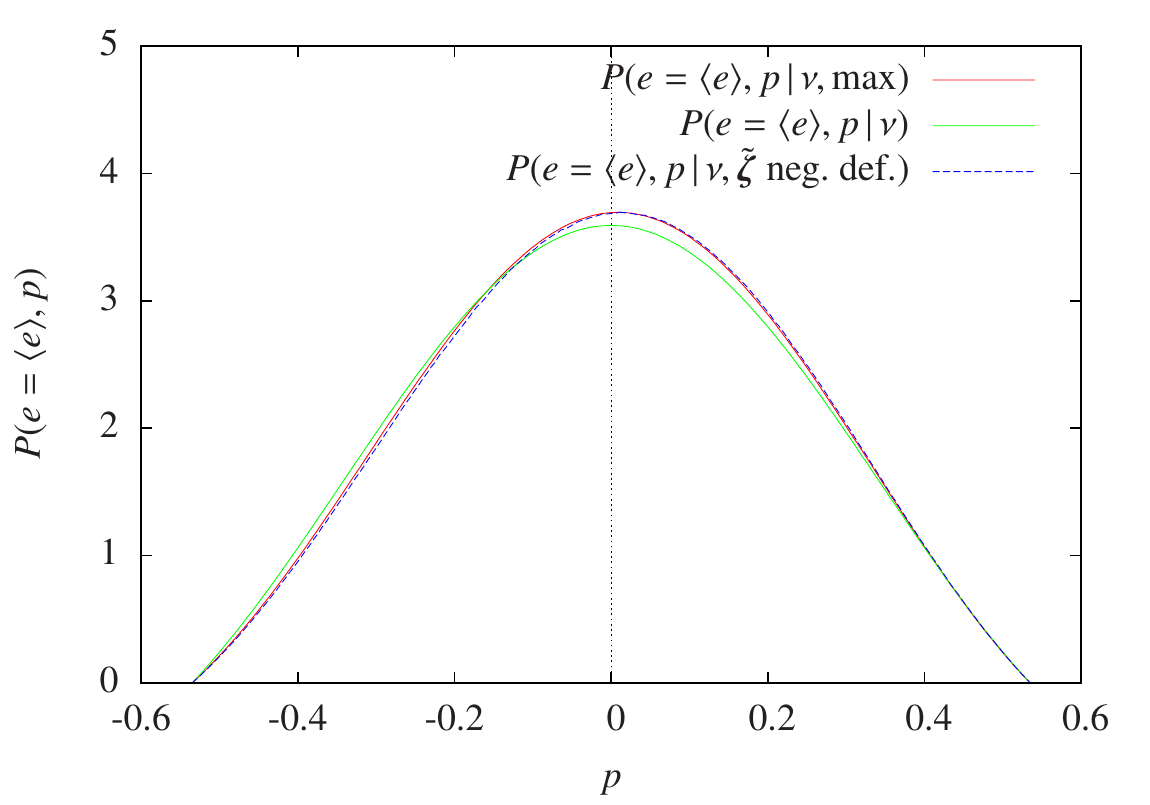}~\includegraphics[width=0.495\textwidth]{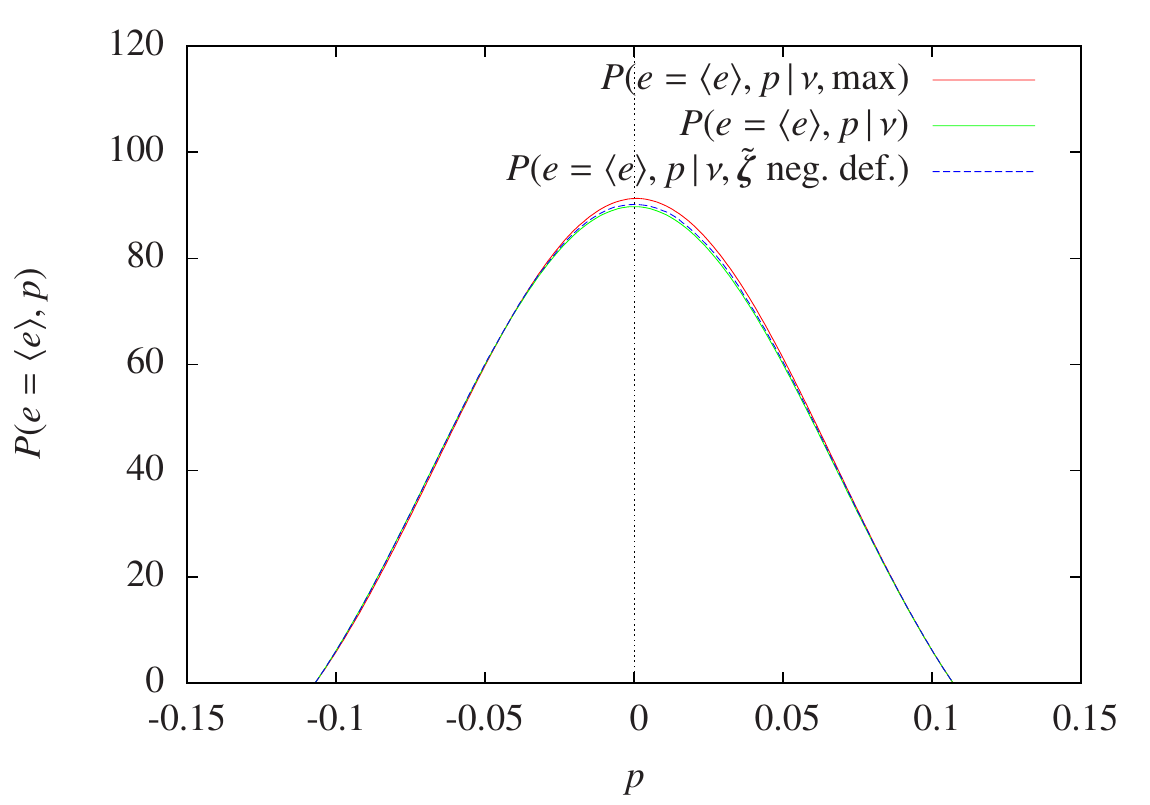}
\caption{Top panels: the distributions $P(e,p\,|\,\nu,\rmn{max})$ based on equation~\eqref{eq:condProb}, $P(e,p\,|\,\nu,\bmath{\tilde{\zeta}}\text{ neg.\ def.})$ based on equation~\eqref{eq:condProbWithout}, and $P(e,p\,|\,\nu)$ based on equation~\eqref{eq:Sheth} for $\nu=1$ (left-hand panels) and $\nu=5$ (right-hand panels). Central panels: same as top panels, but for $p=0$ (cf.\ equation~\ref{eq:analyticExpectationsII}). Bottom panels: same as top panels, but for fixed $e=\langle e\rangle$ (cf.\ equation~\ref{eq:analyticExpectations}).}
\label{fig:2Ddistributions}
\end{figure*}

In this section, I present some numerical results based on the theoretical work developed in the previous section. For the calculation, I adopt a flat reference $\Lambda$CDM model inspired by the \textit{Planck} results \citep{PlanckValues2013} with matter density $\Omega_\rmn{m}=0.315$, dark-energy density $\Omega_\Lambda=0.685$, baryon density $\Omega_\rmn{b}=0.04868$, a Hubble constant of $h=0.673$ in units of $100~\rmn{km}~\rmn{s}^{-1}~\rmn{Mpc}^{-1}$ and power spectrum normalization $\sigma_8\equiv\sigma_0(R=8\ \rmn{Mpc})=0.829$. Note that by this definition, $\sigma_8$ does not correspond to the eighth spectral moment but to the square root of the variance $\sigma_0^2$ filtered on a scale of 8 Mpc. All previous values are set at redshift $z=0$. The primordial spectral index of the power spectrum is set to $n_\rmn{s}=0.9603$, and I adopt the transfer function of \citet{Eisenstein1998}.

\subsection{The joint distribution for ellipticity and prolaticity}
\label{subsecd:jointDistribution}

In the following, I compare the distributions \eqref{eq:condProb} and \eqref{eq:condProbWithout} with the distribution
\begin{equation}
\label{eq:Sheth}
P(e,p\,|\,\tilde{\delta})=\frac{1125}{\sqrt{10\pi}}e(e^2-p^2)\tilde{\delta}^5\exp\left[-\frac{5}{2}\tilde{\delta}^2(3e^2+p^2)\right]
\end{equation}
\citep{Sheth2001}, which does not take any of the two constraints on the first derivative $\bmath{\eta}$ and the Hessian $\bmath{\zeta}$ into account. For a better comparison to other papers in the literature and to simplify the connection to a mass scale, I adopt the variable
\begin{equation}
\label{eq:nu}
\nu\equiv\delta_\rmn{c}/\sigma_0(R)
\end{equation}
instead of $\tilde{\delta}$. In the former equation, $\delta_\rmn{c}$ is the linear overdensity of the spherical-collapse model, and $\sigma_0(R)$ is the power spectrum's variance exponentially filtered on the scale $R$ (cf.\ equation~\ref{eq:specMoments}). Hence, for a given $\nu$, the corresponding reduced density contrast is simply $\tilde{\delta}=\nu$ and the radius $R$ has to be chosen such that equation~\eqref{eq:nu} is fulfilled. It is then used to calculate the higher spectral moments $\sigma_1(R)$ and $\sigma_2(R)$ from equation~\eqref{eq:specMoments} and  also $\gamma(R)=\sigma_1^2(R)/[\sigma_0(R)\,\sigma_2(R)]$.

In Table~\ref{tab:nuMass}, I list the corresponding mass for both $\nu=1$ and $\nu=5$ for some redshifts between 0 and 1 in the cosmological model introduced at the beginning of Section~\ref{sec:results}.

\begin{table}
\caption{Relation between $\nu$ and mass $M$ for a given redshift $z$ assuming the cosmological model of Section~\ref{sec:results}.}
\label{tab:nuMass}
\begin{minipage}{0.243\textwidth}
\centering
\begin{tabular}{cc}
\multicolumn{2}{c}{$\nu=1$} \\
\hline
$z$ & $M\ (h^{-1}\ \rmn{M}_\odot)$ \\[3mm]
0 & $1.95\times10^{12}$ \\
0.2 & $1.00\times10^{12}$ \\
0.4 & $5.01\times10^{11}$ \\
0.6 & $2.47\times10^{11}$ \\
0.8 & $1.22\times10^{11}$ \\
1 & $6.07\times10^{10}$ \\
\end{tabular}
\end{minipage}
\begin{minipage}{0.243\textwidth}
\centering
\begin{tabular}{ccc}
\multicolumn{2}{c}{$\nu=5$} \\
\hline
 $z$ & $M\ (h^{-1}\ \rmn{M}_\odot)$ \\[3mm]
0 & $2.15\times10^{15}$ \\
0.2 & $1.52\times10^{15}$ \\
0.4 & $1.07\times10^{15}$ \\
0.6 & $7.50\times10^{14}$ \\
0.8 & $5.30\times10^{14}$ \\
1 & $3.78\times10^{14}$ \\
\end{tabular}
\end{minipage}
\end{table}

In Fig.~\ref{fig:2Ddistributions}, I compare the distributions \eqref{eq:condProb}, \eqref{eq:condProbWithout}, and \eqref{eq:Sheth} as a function of $e$ and $p$ for two different values of $\nu$. For $\nu=1$, the distributions $P(e,p\,|\,\nu,\rmn{max})$ and $P(e,p\,|\,\nu,\bmath{\tilde{\zeta}}\text{ neg.\ def.})$ (red and blue lines, respectively) coincide very well over the whole range of $e$ and $p$, whereas $P(e,p\,|\,\nu)$ peaks at a slightly larger value of $e$ and as a consequence of the normalization to unity has a smaller amplitude at the maximum compared to the two former distributions. This implies that haloes of lower mass tend to be slightly less elliptical than predicted by the standard formula \eqref{eq:Sheth}.

For $\nu=5$, the situation is reversed: $P(e,p\,|\,\nu,\bmath{\tilde{\zeta}}\text{ neg.\ def.})$ now corresponds very well to $P(e,p\,|\,\nu)$, whereas $P(e,p\,|\,\nu,\rmn{max})$ has a larger amplitude compared to the two former distributions. The amplitude of the relative differences between the distributions, however, has shrunk, indicating that the additional constraints become the less important the higher $\nu$ is.

What is most important for most practical purposes, however, is not the full distribution for $e$ and $p$, but only the values that are either expected or most probable. The expectation values of $e$ and $p$ are used e.g.\ by \citet{Angrick2009} to determine the initial eigenvalues for the ellipsoidal-collapse model by \citet{Bond1996}, whereas the most probable values are used by \citet{Sheth2001} to determine the ellipsoidal barrier when establishing their mass function. The expectation values of $e$ and $p$ as well as their variances can be calculated analytically for the distribution $P(e,p\,|\,\nu)$. They are as a function of $\nu$ given by
\begin{align}
\label{eq:analyticExpectations}
\langle e\rangle&=\frac{3}{\sqrt{10\pi}\nu},&\quad \sigma_e^2&=\frac{19\pi-54}{60\pi\nu^2},\\
\label{eq:analyticExpectationsII}
\langle p\rangle&=0,&\sigma_p^2&=\frac{1}{20\nu^2}
\end{align}
\citep{Angrick2010}. The most probable values for $e$ and $p$ from the same distribution are
\begin{equation}
\label{eq:mostProbable}
e_\rmn{mp}=\frac{1}{\sqrt{5}\nu},\qquad p_\rmn{mp}=0,
\end{equation}
respectively \citep{Sheth2001}.

\begin{figure*}
\centering
\includegraphics[width=0.495\textwidth]{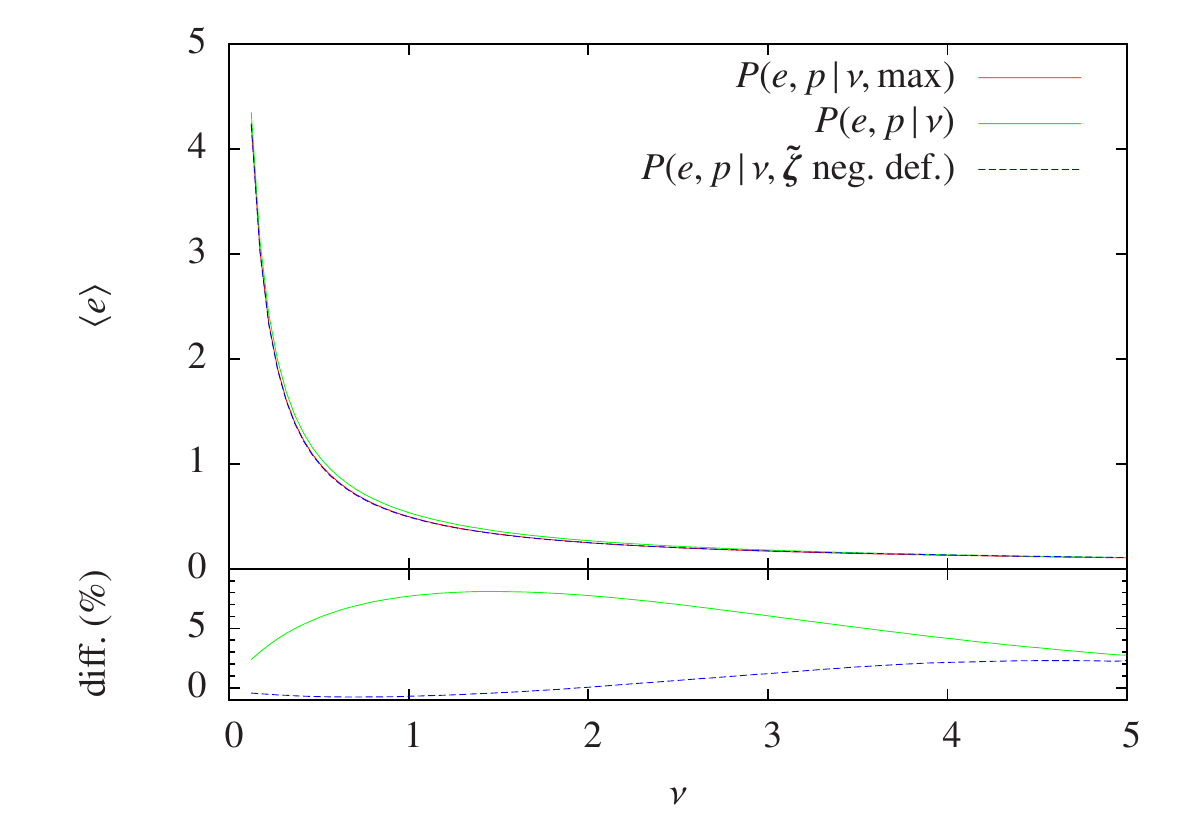}~\includegraphics[width=0.495\textwidth]{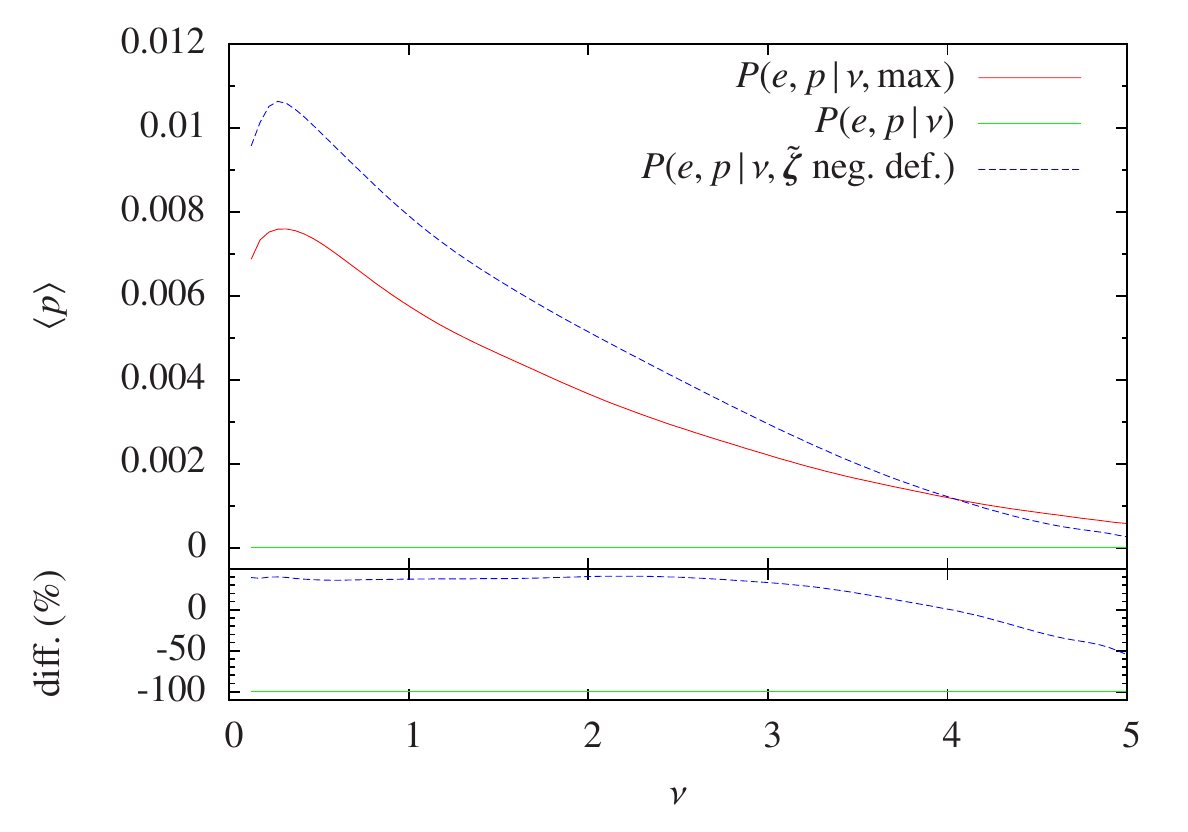}
\caption{Expectation values of the ellipticity (left-hand panel) and the prolaticity (right-hand panel) derived from the distributions $P(e,p\,|\,\nu,\rmn{max})$,  $P(e,p\,|\,\nu,\bmath{\tilde{\zeta}}\text{ neg.\ def.})$, and $P(e,p\,|\,\nu)$ as a function of $\nu$. At the bottom of each plot, the relative differences between the results derived from the latter two distributions with respect to $P(e,p\,|\,\nu,\rmn{max})$ are quantified.}
\label{fig:expectation}
\end{figure*}

In Fig.~\ref{fig:expectation}, I present the expectation values for both ellipticity and prolaticity for the distributions $P(e,p\,|\,\nu,\rmn{max})$, $P(e,p\,|\,\nu,\bmath{\tilde{\zeta}}\text{ neg.\ def.})$, and $P(e,p\,|\,\nu)$ as a function of $\nu$. As expected, the expectation value for the ellipticity is smaller if either \emph{both} constraints for a maximum or only the constraint that the density's Hessian has to be negative definite is taken into account. Additionally, the expected prolaticity acquires a small non-vanishing value in the latter two cases. For larger $\nu$, the differences between the three distributions decrease. While the largest difference in $\langle e\rangle$ is at $\nu\approx 1.4$ with a value that is $\sim$8 per cent larger if none of the maximum constraints are taken into account, $\langle p\rangle$ is monotonically decreasing for $\nu\geq0.2$ with increasing $\nu$ and has a maximal value of $\langle p\rangle\approx0.0075$ at $\nu\approx0.2$ for $P(e,p\,|\,\nu,\rmn{max})$ and $\langle p\rangle\approx0.0105$ for $P(e,p\,|\,\nu,\bmath{\tilde{\zeta}}\text{ neg.\ def.})$ at the same position. Therefore, the most probable value for the prolaticity does not vanish identically for a density maximum as it is the case for arbitrary points.

Interestingly, taking the additional condition of a vanishing first derivative into account does not result in a very different $\langle e\rangle$ (the modulus of the relative difference between the two results is $\lesssim$2 per cent for $\nu\leq 5$ with a change of sign in the deviation at $\nu\approx2$), the relative difference for $\langle p\rangle$ is practically constant with a value that is $\sim$40 per cent larger if the condition on the first derivative is neglected. It decreases until $\langle p\rangle$ is equal for both distributions at $\nu\approx4$, then changes sign and quickly reaches $\sim$50 per cent at $\nu\approx 5$. Although the relative change in the expected prolaticity is large if one incorporates the condition $\bmath{\eta}=\bmath{0}$, the \emph{absolute} amplitudes, however, remain rather small.

Both $\langle e\rangle$ and $\langle p\rangle$ at a maximum reach the values deduced from the joint probability of $e$ and $p$ not taking any of the two maximum constraints into account as $\nu\rightarrow\infty$ for the same reason as already stated above: The higher $\nu$ is, the more probable it is that the maximum constraint is automatically fulfilled.

The same quantitative behaviour is found for the most probable values $e_\rmn{mp}$ and $p_\rmn{mp}$. The deviations are $\lesssim$8 per cent for $e_\rmn{mp}$ and $\lesssim$50 per cent for $p_\rmn{mp}$ with the same dependence on $\nu$ as for the expectation values $\langle e\rangle$ and $\langle p\rangle$ so that the latter approaches zero for increasing $\nu$.

In summary, the ellipticity used in analytical models of structure formation is about 3--8 per cent too high in the range of $\nu$ relevant for cosmology since the maximum constraint is not taken into account. The prolaticity is usually assumed to vanish, but should be slightly positive with values at the level of $10^{-3}$--$10^{-2}$ in the same range.

\subsection{Implications for the ellipsoidal-collapse model}
\label{subsec:ellModel}

In this section, I will quantify the consequences for the ellipsoidal-collapse model by \citet{Bond1996}, extended by \citet{Angrick2010}, especially for the parameters $\delta_\rmn{c}$ and $\Delta_\rmn{v}$. Before, I will present the most important formulae.

The evolution of the scaled dimensionless principal axes $a_i=R_i/R_\rmn{pk}$ with $1\leq i\leq 3$ of a homogeneous ellipsoid, where the $R_i$ are the dimensional axes and $R_\rmn{pk}$ is the radius of a sphere with homogeneous density $\rho_\rmn{b}$ that contains a mass $M$, is governed by the following three coupled differential equations:
\begin{equation}
 \label{eq:basicEvolutionA}
\frac{\dd^2a_i}{\dd a^2}+\left[\frac{1}{a}+\frac{E'(a)}{E(a)}\right]\frac{\dd a_i}{\dd a}+\left[\frac{3\Omega_\rmn{m}}{2a^5 E^2(a)}C_i(a)-\frac{\Omega_\Lambda}{a^2 E^2(a)}\right]a_i=0,
\end{equation}
where a prime denotes differentiation with respect to the scale factor $a$, $E(a)$ is the universe's expansion function, and $C_i\equiv(1+\delta_\rmn{ell})/3+b_i/2+\lambda_{\rmn{ext},i}$. Here, $\delta_\rmn{ell}\equiv a^3/(a_1 a_2 a_3)-1$ is the density contrast inside the ellipsoid, and the $\lambda_{\rmn{ext},i}$ are the external shear's eigenvalues whose evolution is modelled according to the \emph{hybrid model} so that
\begin{equation}
\label{eq:defineExtShear}
\lambda_{\rmn{ext},i}(a)\equiv
\begin{cases}
\dfrac{5}{4}b_i(a) &\quad\rmn{if}~a\leq a_{\rmn{ta,i}}, \\[3mm]
\dfrac{D_+(a)}{D_+(a_{\rmn{ta},i})}\lambda_{\rmn{ext},i}(a_{\rmn{ta},i}) &\quad\rmn{else},
\end{cases}
\end{equation}
where $a_{\rmn{ta},i}$ is the turn-around scale factor of the $i$th axis, $D_+(a)$ is the linear growth factor of structure formation, and the $b_i$ are the internal-shear tensor's eigenvalues given by
\begin{equation}
 \label{eq:defineIntShear}
b_i(a)\equiv a_1(a)\,a_2(a)\,a_3(a)\int_0^\infty\frac{\dd\tau}{[a_i^2(a)+1]\prod_{k=1}^3[a_k^2(a)+1]^{1/2}}-\frac{2}{3}.
\end{equation}
Note that in this type of ellipsoidal-collapse model, the ellipsoid's principal axes are aligned with those of the external gravitational shear.

\begin{figure*}
\centering
\includegraphics[width=0.495\textwidth]{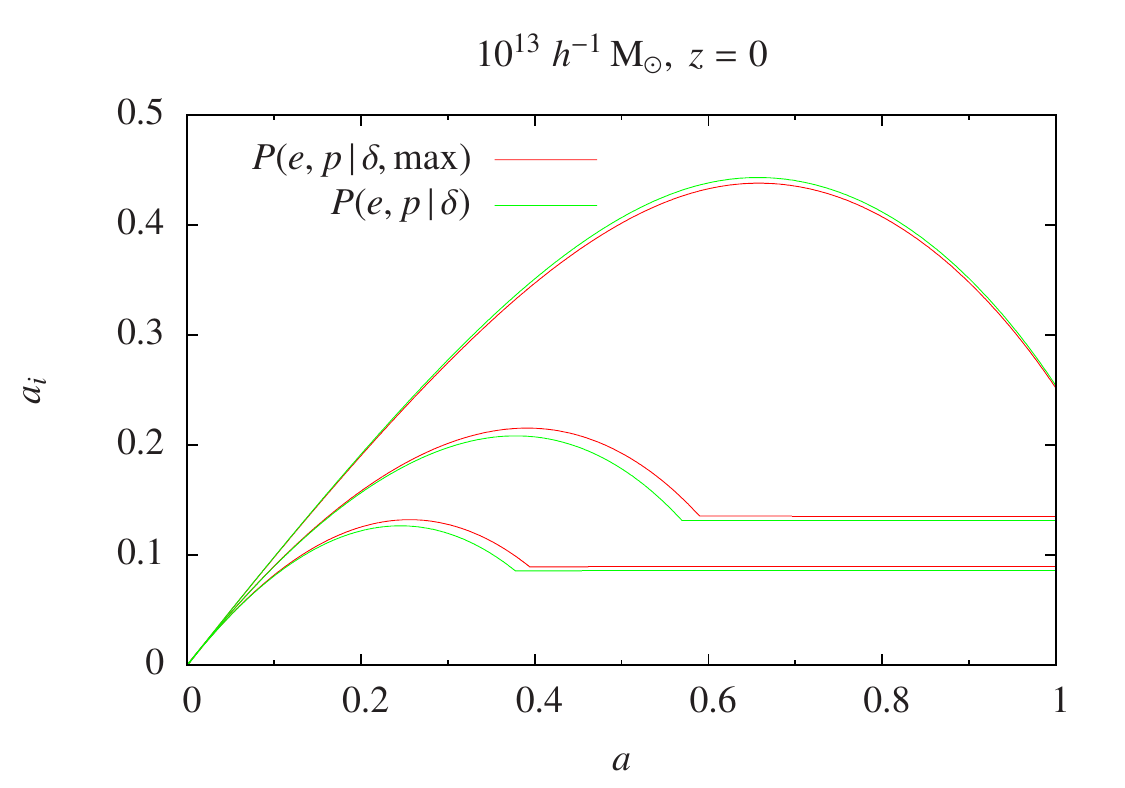}~\includegraphics[width=0.495\textwidth]{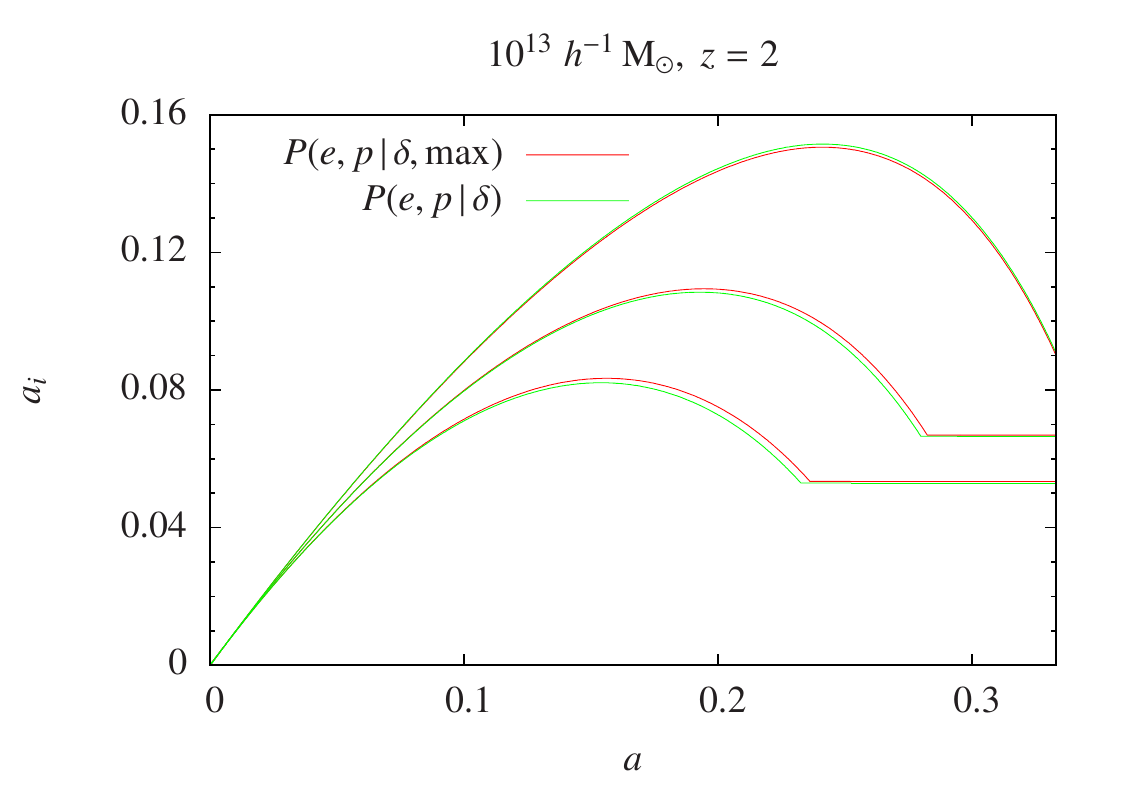}\\[2mm]
\includegraphics[width=0.495\textwidth]{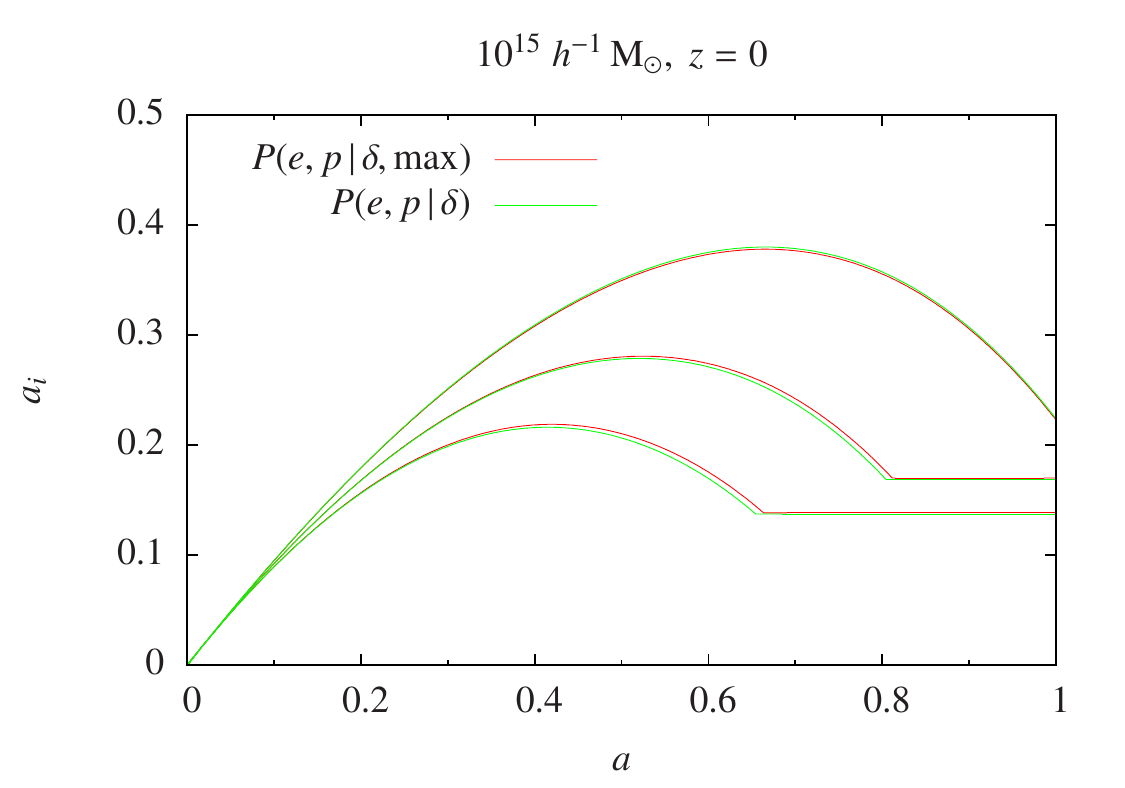}~\includegraphics[width=0.495\textwidth]{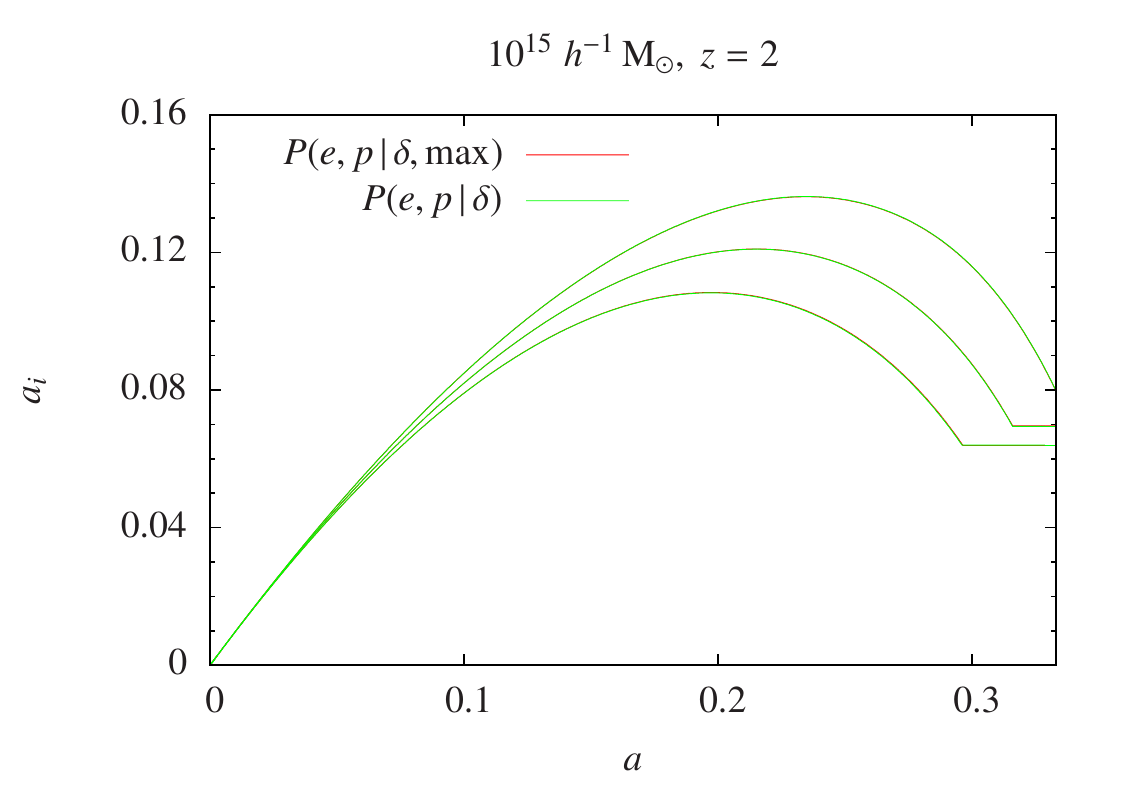}
\caption{Evolution of an ellipsoid's axes using either $P(e,p\,|\,\delta)$ or $P(e,p\,|\,\delta,\rmn{max})$ when determining the axes' initial conditions via equations~\eqref{eq:initAxes}--\eqref{eq:lambda3} for two masses ($10^{13}\ h^{-1} \rmn{M}_\odot$ in the upper panels and $10^{15}\ h^{-1} \rmn{M}_\odot$ in the lower panels) and two virialization redshifts (0 in the left-hand panels and 2 in the right-hand panels).}
\label{fig:axes}
\end{figure*}

The initial conditions are chosen such that they are compatible with the Zel'dovich approximation at early times. Hence,
\begin{align}
 \label{eq:initAxes}
a_i(a_0)&=a_0[1-\lambda_i(a_0)], \\
\label{eq:initAxesVel}
\left.\frac{\dd a_i}{\dd a}\right|_{a_0}&=1-\lambda_i(a_0)-\left.\frac{\dd\ln D_+}{\dd\ln a}\right|_{a_0}\lambda_i(a_0)\approx 1-2\lambda_i(a_0)
\end{align}
since $D_+(a)\approx a$ for $a\ll1$ which is safely fulfilled for $a_0=2\times10^{-5}$. At this point, a change in the joint distribution for the ellipticity and the prolaticity has influence on the model's results since the $\lambda_i(a_0)$ are the initial eigenvalues of the gravitational-shear tensor, whose elements were defined in equation~\eqref{eq:deformationTensor}. The eigenvalues are functions of $\langle e\rangle$, $\langle p\rangle$, and $\delta$ up to first order in $e$ and $p$,
\begin{align}
\label{eq:lambda1}
\lambda_1&=\frac{\delta}{3}(1+3\langle e\rangle+\langle p\rangle),\\
\label{eq:lambda2}
\lambda_2&=\frac{\delta}{3}(1-2\langle p\rangle),\\
\label{eq:lambda3}
\lambda_3&=\frac{\delta}{3}(1-3\langle e\rangle +\langle p\rangle),
\end{align}
implying $\lambda_1\geq\lambda_2\geq\lambda_3$.

The collapse of the $i$th axis is stopped when the following virialization conditions derived from the tensor virial theorem are fulfilled,
\begin{equation}
 \label{eq:virCondition}
\left(\frac{a_i'}{a_i}\right)^2=\frac{1}{a^2 E^2(a)}\left(\frac{3\Omega_\rmn{m}}{2a^3}C_i-\Omega_\Lambda\right)\quad\rmn{and}\quad a_i'<0.
\end{equation}

The parameters $\delta_\rmn{c}$ and $\Delta_\rmn{v}$ can be calculated from the ellipsoidal-collapse model by
\begin{equation}
 \label{eq:defineParam}
\delta_\rmn{c}=\frac{D_+(a_\rmn{v})}{D_+(a_0)}\sum_{i=1}^3\lambda_i(a_0),\qquad\Delta_\rmn{v}=\frac{\Omega_\rmn{m}(a_\rmn{v})\,a_\rmn{v}^3}{a_1(a_\rmn{v})\,a_2(a_\rmn{v})\,a_3(a_\rmn{v})},
\end{equation} 
where $a_\rmn{v}$ is the scale factor at which the third axis virializes.

In Fig.~\ref{fig:axes}, I present the evolution of the ellipsoid's three axes as a function of the scale factor for two different virialization redshifts and two masses using either $P(e,p\,|\,\delta)$ or $P(e,p\,|\,\delta,\rmn{max})$ for the determination of the axes' initial conditions. These distributions enter the model via equations~\eqref{eq:initAxes} and \eqref{eq:initAxesVel} since both $a_i$ and $\dot{a}_i$ are dependent on the respective gravitational-shear tensor's eigenvalue which is a linear combination of $\langle e\rangle$ and $\langle p\rangle$. For clarity, I did not include the evolution of the ellipsoid's axes for which the initial conditions are based on $P(e,p\,|\,\delta,\bmath{\tilde{\zeta}}\text{ neg.\ def.})$ since the differences to the results for $P(e,p\,|\,\delta,\rmn{max})$ are negligible.

All four plots have in common that the two axes along which virialization sets in first ($a_1$ and $a_2$) are \emph{smaller} if the maximum constraint is not taken into account, whereas $a_3$, the longest axis, is \emph{larger} in comparison. Additionally, the virialization of $a_1$ and $a_2$ sets in earlier. These features are most pronounced for small masses and low redshifts.

The reason is as follows. For $P(e,p\,|\,\delta,\rmn{max})$, $\langle e\rangle$ is lowered, whereas $\langle p\rangle$ is enlarged according to Fig.~\ref{fig:expectation}, but in terms of absolute values this effect is larger for the ellipticity than for the prolaticity. Hence, equations~\eqref{eq:lambda1}--\eqref{eq:lambda3} imply that both $\lambda_1$ and $\lambda_2$ are lowered, whereas $\lambda_3$ is enlarged so that according to equations~\eqref{eq:initAxes} and \eqref{eq:initAxesVel}, the axes $a_1$ and $a_2$ both start with slightly larger amplitudes and velocities, and $a_3$ is slightly smaller in both amplitude and velocity.

\begin{figure*}
\centering
\includegraphics[width=0.495\textwidth]{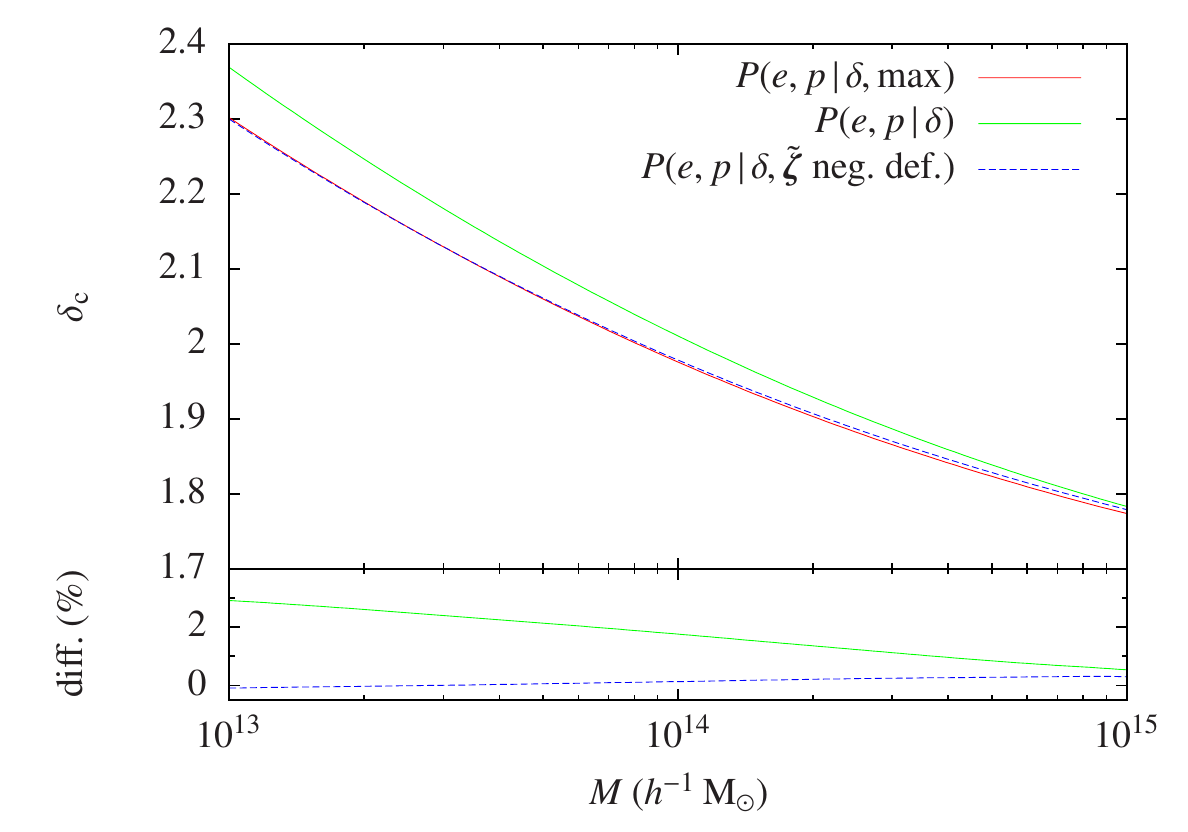}~\includegraphics[width=0.495\textwidth]{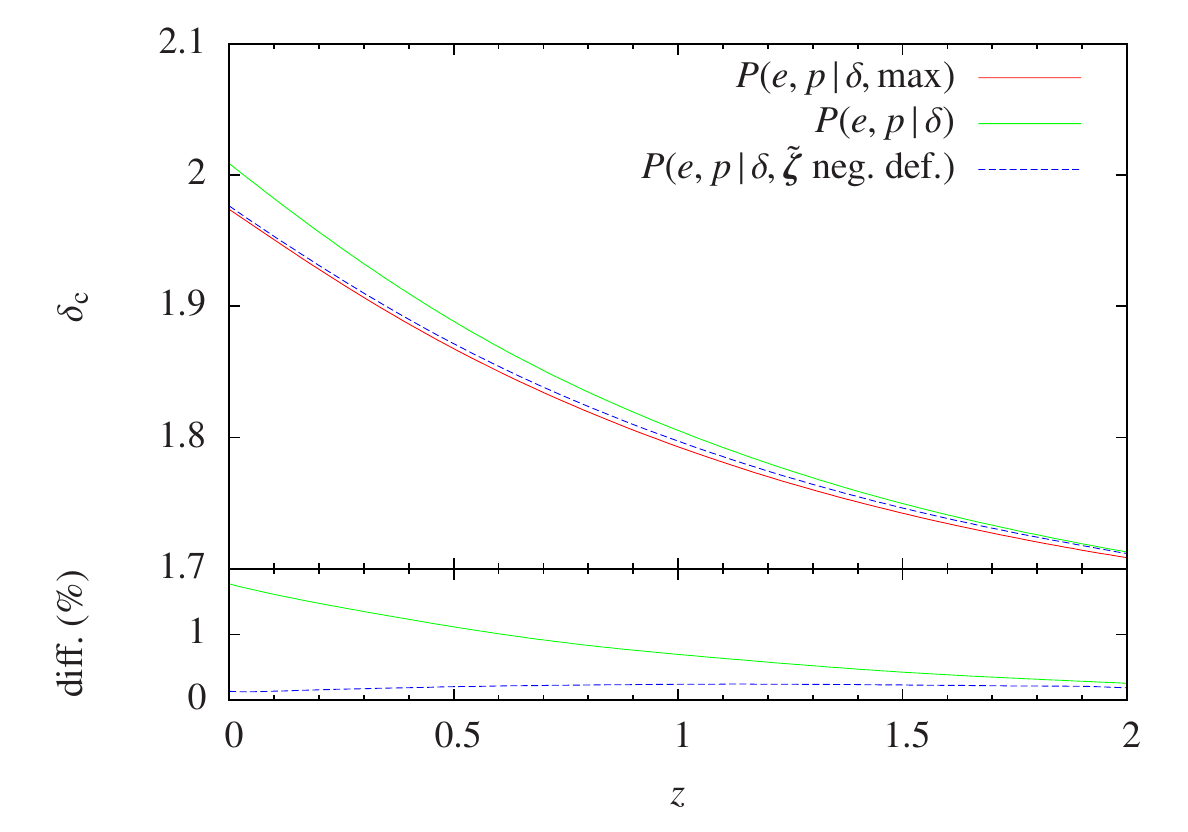}\\[2mm]
\includegraphics[width=0.495\textwidth]{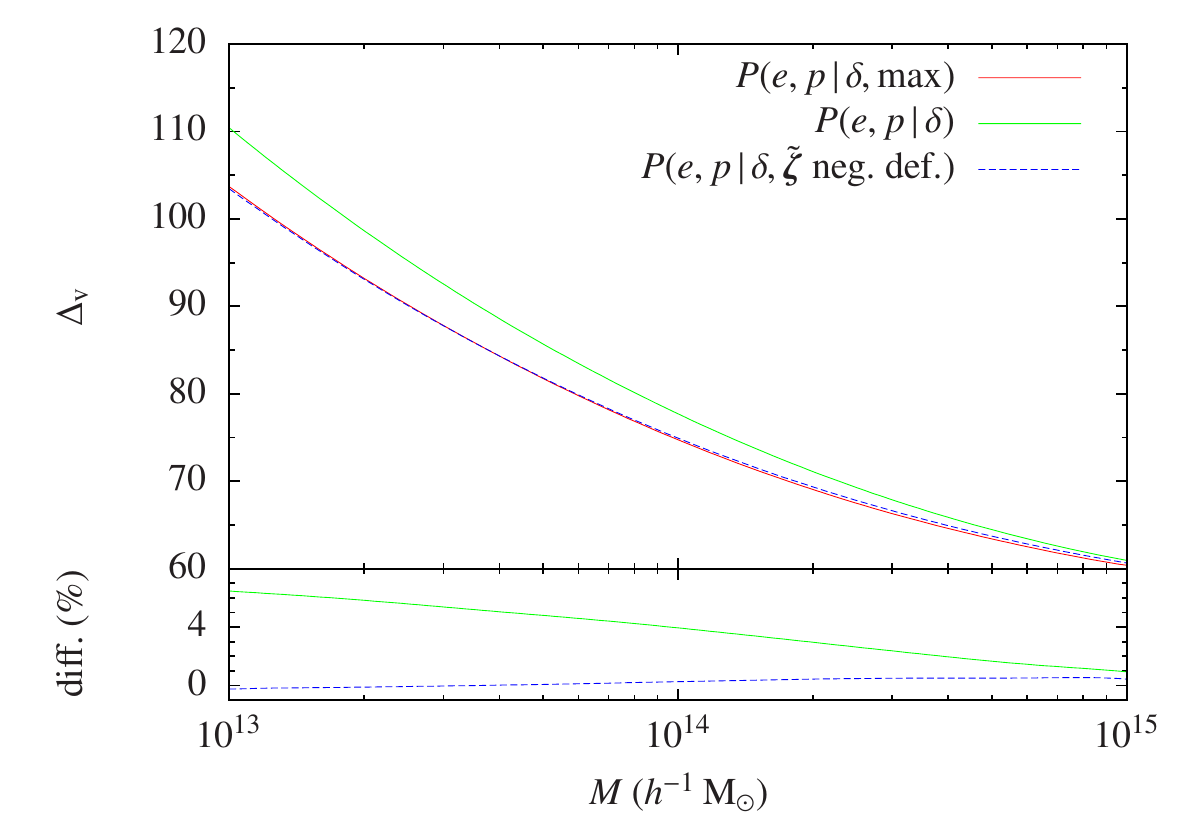}~\includegraphics[width=0.495\textwidth]{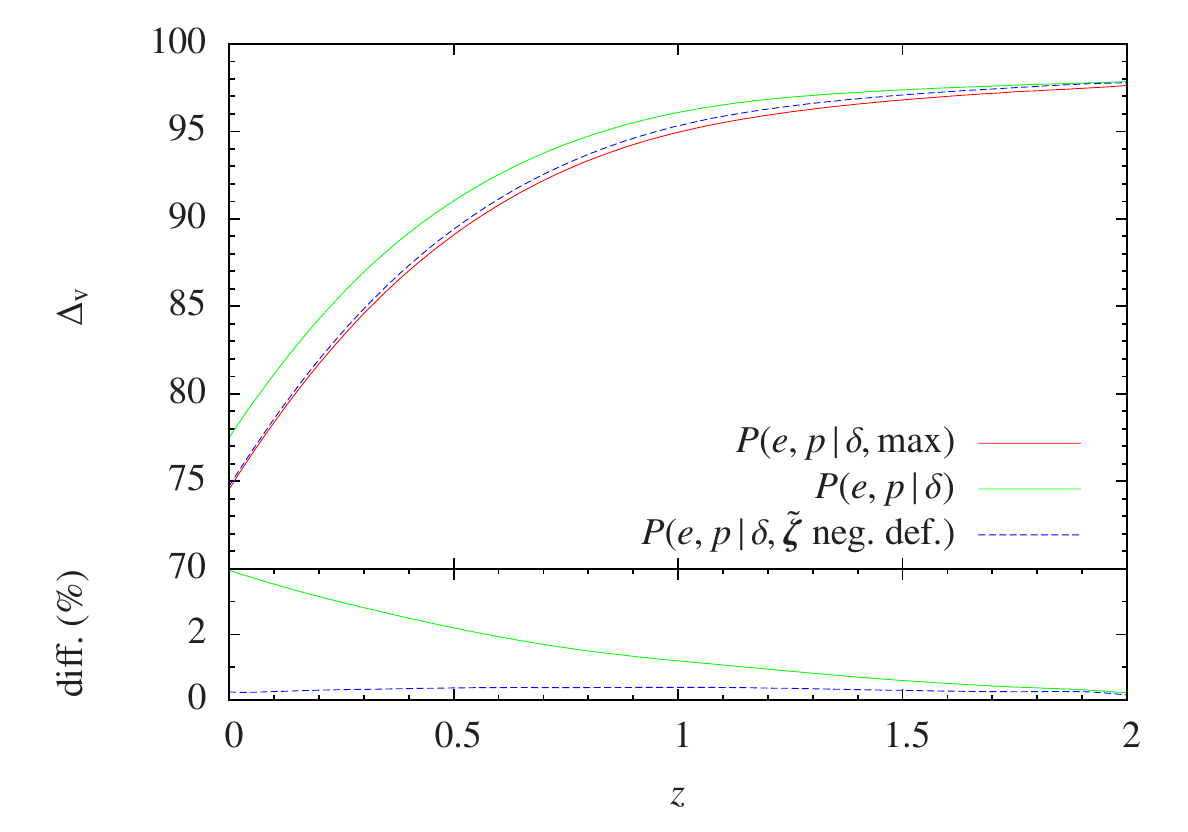}
\caption{The parameters $\delta_\rmn{c}$ and $\Delta_\rmn{v}$ as a function of mass at redshift zero (left-hand panels) and as a function of redshift at a fixed mass of $10^{14}~h^{-1}~\rmn{M}_\odot$ (right-hand panels).}
\label{fig:deltas}
\end{figure*}

For large masses and high redshifts, however, the ellipsoid's evolution differs less since a larger overdensity is needed for the ellipsoid to collapse at the redshift considered so that the change in $\langle e\rangle$ and $\langle p\rangle$ is much less pronounced as shown in Fig.~\ref{fig:expectation}. Hence, the smallest difference in the evolution can be found in Fig.~\ref{fig:axes} for the ellipsoid with $10^{15}~h^{-1}~\rmn{M}_\odot$ virializing at $z=2$. Additionally, the collapse becomes less ellipsoidal at larger masses and higher redshifts.

The results of this change in the axes' evolution for the parameters $\delta_\rmn{c}$ and $\Delta_\rmn{v}$ are shown in Fig.~\ref{fig:deltas}. The overall trend is that both parameters are \emph{lowered} if additional constraints on the density's spatial derivatives are taken into account. This effect is the stronger the lower the ellipsoid's virialization redshift and its mass are. Additionally to the constraint that the density's Hessian is negative definite, incorporating the second constraint of a vanishing first derivative changes both $\delta_\rmn{c}$ and $\Delta_\rmn{v}$ by less than half a per cent over the whole mass and redshift ranges considered.

For ellipsoids with a mass $M$ between $10^{13}$ and $10^{15}~h^{-1}~\rmn{M}_\odot$ and a virialization redshift $z$ between 0 and 2, $\delta_\rmn{c}$ is $\lesssim$3 per cent smaller if the maximum constraints are not accounted for, where the maximal deviation of $\sim$3 per cent is reached for $M=10^{13}~h^{-1}~\rmn{M}_\odot$ and $z=0$.  The deviation drops to $\sim$2 per cent for $M=10^{14}~h^{-1}~\rmn{M}_\odot$ and $\sim$0.5 per cent for $10^{15}~h^{-1}~\rmn{M}_\odot$ both at $z=0$.

The effect for $\Delta_\rmn{v}$ is approximately twice as large as for $\delta_\rmn{c}$ with values for $\Delta_\rmn{v}$ that are $\lesssim$7.5 per cent smaller if $P(e,p\,|\,\delta)$ instead of $P(e,p\,|\,\delta,\rmn{max})$ is used for the initial conditions. Again, the largest deviation can be found for $M=10^{13}~h^{-1}~\rmn{M}_\odot$ and $z=0$. The difference drops to $\sim$4 per cent for $M=10^{14}~h^{-1}~\rmn{M}_\odot$ and $\sim$1 per cent for $M=10^{15}~h^{-1}~\rmn{M}_\odot$ both at $z=0$.

While only the sum $\sum_{i=1}^3\lambda_i=\delta$ enters the computation of $\delta_\rmn{c}$, whose evolution is then driven by the linear growth factor $D_+(a)$, the situation is different for $\Delta_\rmn{v}$. Here, the individual axes $a_i$ are initially altered due to a change of initial $\delta$, $\langle e\rangle$ and $\langle p\rangle$ and evolve non-linearly. The $a_i$ at the time of virialization then determine $\Delta_\rmn{v}$ according to equation~\eqref{eq:defineParam}. This non-linearity is the reason why the deviation for $\Delta_\rmn{v}$ is larger than for $\delta_\rmn{c}$. 

\section{Summary \& conclusions}
\label{sec:summary}

In the first part of this article, I have derived the probability for the gravitational-shear field's ellipticity and prolaticity at the position of a peak in the density field with a given height. The derivation is based on the statistics of Gaussian random fields \citep{Bardeen1986} and extends the works by \citet{Sheth2001} and \citet{Rossi2012,Rossi2013} as follows. In contrast to \citet{Sheth2001}, I have added explicitly the maximum constraint in the calculation since it is expected that haloes form at the positions of peaks in the density field. Although \citet{Rossi2012,Rossi2013} started going into the same direction, they have considered the conditional eigenvalues of the matrix $\tilde{\bmath{\varphi}}\,|\,\tilde{\bmath{\zeta}}$ and the corresponding ellipticity and prolaticity. These, however, do not serve as ingredients for the ellipsoidal-collapse model by \citet{Bond1996} since they use the ellipticity and the prolaticity derived from the eigenvalues of the matrix $\tilde{\bmath{\varphi}}$ instead. Additionally, they considered that the density's Hessian has to be negative definite for a maximum but did not take into account that the first derivative $\bmath{\eta}$ also has to vanish.

The main results of the first part can be summarized as follows:
\begin{itemize}
\item The gravitational-shear tensor $\bmath{\varphi}$ and the density's Hessian $\bmath{\zeta}$ cannot be diagonalized simultaneously. Since the eigenvalues of $\bmath{\varphi}$ are needed for the definition of the ellipticity $e$ and prolaticity $p$ as used in the ellipsoidal-collapse model by \citet{Bond1996}, the priority is on the diagonalisation of $\bmath{\varphi}$ so that first, the constraint that the density's Hessian $\bmath{\zeta}$ has to be negative definite for a maximum can only be considered via Sylvester's criterion and second, the full six-dimensional integration over $\dd^6\zeta$ has to be carried out.
\item The conditional probability for $e$ and $p$ given a density maximum of height $\delta$ including the constraints on \emph{both} the first and second derivatives is given by \eqref{eq:condProb} after rescaling to $\tilde{\delta}=\delta/\sigma_0$. The numerator can be calculated by integrating equation~\eqref{eq:jointProb} multiplied by the function \eqref{eq:f} that takes into account Sylvester's criterion for a negative definite Hessian of the density, while the denominator is given by equation~\eqref{eq:numDensMax} divided by $\sigma_2^3/\sigma_1^3$ together with equation~\eqref{eq:numDensSubs}.
\item The conditional probability for $e$ and $p$ only incorporating the constraint that the density's Hessian has to be negative definite is given by \eqref{eq:condProbWithout} in rescaled quantities, where the numerator can be calculated using equation~\eqref{eq:numeratorWithout}, while the denominator is given by equation~\eqref{eq:downIntegrate} together with equation~\eqref{eq:downIntegrateSubs}.
\item The distribution of $e$ and $p$ incorporating the maximum constraint peaks at lower ellipticities and slightly positive prolaticities compared to the unconstrained one by \citet{Sheth2001}.
\item As a consequence, both the expected and the most probable ellipticities are about 3--8 per cent larger in the range $1\leq\nu\leq 5$ if the maximum constraint is not accounted for with the largest deviation at around $\nu\sim1.4$. Both the expected and the most probable prolaticity have slightly positive values at the level of $10^{-3}$--$10^{-2}$ reaching the largest absolute value at $\nu\sim0.2$.
\item Both expected and most probable values for ellipticity and prolaticity from the distribution $P(e,p\,|\,\nu,\rmn{max})$ tend to converge to the values from the distribution $P(e,p\,|\,\nu)$ for $\nu\rightarrow\infty$ since the larger $\nu$, the more likely is the maximum constraint automatically fulfilled.
\end{itemize}

In the second part of this article, I quantify the impact on the ellipsoidal-collapse model by \citet{Bond1996} that was slightly modified and extended by \citet{Angrick2010} if either both maximum constraints or only the constraint on the density's Hessian are/is taken into account, especially on the parameters $\delta_\rmn{c}$ and $\Delta_\rmn{v}$. The main results of the second part are as follows.

\begin{itemize}
\item If none of the additional constraints is taken into account, the resulting linear overdensity $\delta_\rmn{c}$ is up to 4 per cent higher compared to the constrained distribution $P(e,p\,|\,\nu,\rmn{max})$, and the virial overdensity $\Delta_\rmn{v}$ is up to 6 per cent higher in the mass range of $10^{13}$--$10^{15}~h^{-1}~\rmn{M}_\odot$ and in the redshift range of 0--2. The deviations between the results from the two distributions decrease for both increasing mass and redshift.
\item The results on the parameters $\delta_\rmn{c}$ and $\Delta_\rmn{v}$ differ less than 0.5 per cent if the constraint that the first derivative has to vanish is neglected.
\end{itemize}

Although the calculation of $\langle e\rangle$ and $\langle p\rangle$ at the position of density maxima cannot be carried out analytically any more, and it is computationally rather expensive ($\sim$1 min on an \emph{Intel Pentium(R)} with 2.80 GHz) due to an eight-dimensional integral that has to be carried out, the final results for both $\delta_\rmn{c}$ and $\Delta_\rmn{v}$ only change mildly. Therefore, it depends on the individual application of the ellipsoidal-collapse model if the gain of accuracy of a few percent in both parameters is more important than the long computation time that can be saved if the maximum constraint is neglected.

\section*{Acknowledgements}
I want to thank M.~Bartelmann for carefully reading the manuscript and for the suggestions that helped to improve it and B.\,M.~Sch\"afer for the discussions that finally led to the implementation of Sylvester's criterion.

\bibliography{bibliography.bib}

\end{document}